\documentclass{aastex63}

\expandafter\ifx\csname natexlab\endcsname\relax\def\natexlab#1{#1}\fi
\providecommand{\url}[1]{\href{#1}{#1}}
\providecommand{\dodoi}[1]{doi:~\href{http://doi.org/#1}{\nolinkurl{#1}}}
\providecommand{\doeprint}[1]{\href{http://ascl.net/#1}{\nolinkurl{http://ascl.net/#1}}}
\providecommand{\doarXiv}[1]{\href{https://arxiv.org/abs/#1}{\nolinkurl{https://arxiv.org/abs/#1}}}

\submitjournal{AJ}

\shorttitle{Mini-EUSO on the ISS}
\shortauthors{Bacholle et al.}

\begin{document}

\title{Mini-EUSO mission to study Earth UV emissions on board the ISS}

\correspondingauthor{Marco Casolino}
\email{marco.casolino@roma2.infn.it}

\correspondingauthor{Laura Marcelli}
\email{laura.marcelli@roma2.infn.it}

\author{S. Bacholle}
\affiliation{Universit\'e de Paris, CNRS, Astroparticule et Cosmologie, F-75006 Paris, France}

\author{P. Barrillon}
\affiliation{LAL, Univ. Paris-Sud, CNRS/IN2P3, Universit\'e Paris-Saclay, Orsay, France}

\author{M. Battisti}
\affiliation{Istituto Nazionale di Fisica Nucleare - Sezione di Torino, Italy}
\affiliation{Dipartimento di Fisica, Universit\'a di Torino, Italy}

\author{A. Belov}
\affiliation{Faculty of physics, M.V. Lomonosov Moscow State University, Moscow, Russia}
\affiliation{Skobeltsyn Institute of Nuclear Physics (SINP), M.V. Lomonosov Moscow State University, Moscow,
Russia}

\author{M. Bertaina}
\affiliation{Istituto Nazionale di Fisica Nucleare - Sezione di Torino, Italy}
\affiliation{Dipartimento di Fisica, Universit\'a di Torino, Italy}

\author{F. Bisconti}
\affiliation{Istituto Nazionale di Fisica Nucleare - Sezione di Torino, Italy}
\affiliation{Dipartimento di Fisica, Universit\'a di Torino, Italy}

\author{C. Blaksley}
\affiliation{RIKEN, Wako, Japan}

\author{S. Blin-Bondil}
\affiliation{Omega, Ecole Polytechnique, CNRS/IN2P3, Palaiseau, France}

\author{F. Cafagna}
\affiliation{Istituto Nazionale di Fisica Nucleare - Sezione di Bari, Italy}

\author{G. Cambi\`e}
\affiliation{Istituto Nazionale di Fisica Nucleare - Sezione di Roma Tor Vergata, Italy}
\affiliation{Universit\'a degli Studi di Roma Tor Vergata - Dipartimento di Fisica, Roma, Italy}

\author{F. Capel}
\affiliation{KTH Royal Institute of Technology, Stockholm, Sweden}

\author{M. Casolino}
\affiliation{Istituto Nazionale di Fisica Nucleare - Sezione di Roma Tor Vergata, Italy}
\affiliation{RIKEN, Wako, Japan}
\affiliation{Universit\'a degli Studi di Roma Tor Vergata - Dipartimento di Fisica, Roma, Italy}

\author{M. Crisconio}
\affiliation{ASI, Italian Space Agency, Italy}

\author{I. Churilo}
\affiliation{S.P. Korolev Rocket and Space Corporation Energia, Korolev, Moscow area, Russia}

\author{G. Cotto}
\affiliation{Istituto Nazionale di Fisica Nucleare - Sezione di Torino, Italy}
\affiliation{Dipartimento di Fisica, Universit\'a  di Torino, Italy}

\author{C. de la Taille }
\affiliation{Omega, Ecole Polytechnique, CNRS/IN2P3, Palaiseau, France}

\author{A. Djakonow}
\affiliation{National Centre for Nuclear Research, Lodz, Poland}

\author{T. Ebisuzaki}
\affiliation{RIKEN, Wako, Japan}

\author{F. Fenu}
\affiliation{Istituto Nazionale di Fisica Nucleare - Sezione di Torino, Italy}
\affiliation{Istituto Nazionale di Astrofisica - Osservatorio astrofisico di Torino, Italy}

\author{A. Franceschi}
\affiliation{Istituto Nazionale di Fisica Nucleare - Laboratori Nazionali di Frascati, Italy}

\author{C. Fuglesang}
\affiliation{KTH Royal Institute of Technology, Stockholm, Sweden}

\author{P. Gorodetzky}
\affiliation{APC, Univ Paris Diderot, CNRS/IN2P3, CEA/Irfu, Obs de Paris, Sorbonne Paris Cit\'e, France}

\author{A. Haungs}
\affiliation{Karlsruhe Institute of Technology, Karlsruhe, Germany}

\author{F. Kajino}
\affiliation{Konan University, Kobe, Japan}

\author{H. Kasuga}
\affiliation{RIKEN, Wako, Japan}

\author{B. Khrenov}
\affiliation{Skobeltsyn Institute of Nuclear Physics (SINP), M.V. Lomonosov Moscow State University, Moscow,
Russia}

\author{P. Klimov}
\affiliation{Skobeltsyn Institute of Nuclear Physics (SINP), M.V. Lomonosov Moscow State University, Moscow,
Russia}

\author{S. Kochepasov}
\affiliation{Skobeltsyn Institute of Nuclear Physics (SINP), M.V. Lomonosov Moscow State University, Moscow,
Russia}

\author{V. Kuznetsov}
\affiliation{S.P. Korolev Rocket and Space Corporation Energia, Korolev, Moscow area, Russia}

\author{L. Marcelli}
\affiliation{Istituto Nazionale di Fisica Nucleare - Sezione di Roma Tor Vergata, Italy}

\author{W. Marsza{\l}}
\affiliation{National Centre for Nuclear Research, Lodz, Poland}

\author{M. Mignone}
\affiliation{Istituto Nazionale di Fisica Nucleare - Sezione di Torino, Italy}

\author{G. Mascetti}
\affiliation{ASI, Italian Space Agency, Italy}

\author{H. Miyamoto}
\affiliation{Istituto Nazionale di Fisica Nucleare - Sezione di Torino, Italy}
\affiliation{Dipartimento di Fisica, Universit\'a  di Torino, Italy}

\author{A. Murashov}
\affiliation{Skobeltsyn Institute of Nuclear Physics (SINP), M.V. Lomonosov Moscow State University, Moscow,
Russia}

\author{T. Napolitano}
\affiliation{Istituto Nazionale di Fisica Nucleare - Laboratori Nazionali di Frascati, Italy}

\author{A.~V.~Olinto}
\affiliation{The University of Chicago, Chicago, IL, USA}

\author{H. Ohmori}
\affiliation{RIKEN, Wako, Japan}

\author{G. Osteria}
\affiliation{Istituto Nazionale di Fisica Nucleare - Sezione di Napoli, Italy}

\author{M. Panasyuk}
\affiliation{Faculty of physics, M.V. Lomonosov Moscow State University, Moscow, Russia}
\affiliation{Skobeltsyn Institute of Nuclear Physics (SINP), M.V. Lomonosov Moscow State University, Moscow,
Russia}

\author{M. Porfilio}
\affiliation{ASI, Italian Space Agency, Italy}

\author{A. Poroshin}
\affiliation{Skobeltsyn Institute of Nuclear Physics (SINP), M.V. Lomonosov Moscow State University, Moscow,
Russia}

\author{E. Parizot}
\affiliation{Universit\'e de Paris, CNRS, Astroparticule et Cosmologie, F-75006 Paris, France}

\author{P. Picozza}
\affiliation{Universit\'a degli Studi di Roma Tor Vergata - Dipartimento di Fisica, Roma, Italy}
\affiliation{Istituto Nazionale di Fisica Nucleare - Sezione di Roma Tor Vergata, Italy}

\author{L. W. Piotrowski}
\affiliation{RIKEN, Wako, Japan}

\author{Z. Plebaniak}
\affiliation{Istituto Nazionale di Fisica Nucleare - Sezione di Torino, Italy}
\affiliation{Dipartimento di Fisica, Universit\'a  di Torino, Italy}
\affiliation{National Centre for Nuclear Research, Lodz, Poland}

\author{G. Pr\'ev\^ot}
\affiliation{Universit\'e de Paris, CNRS, Astroparticule et Cosmologie, F-75006 Paris, France}

\author{M. Przybylak}
\affiliation{National Centre for Nuclear Research, Lodz, Poland}

\author{E. Reali}
\affiliation{Universit\'a degli Studi di Roma Tor Vergata - Dipartimento di Fisica, Roma, Italy}

\author{M. Ricci}
\affiliation{Istituto Nazionale di Fisica Nucleare - Laboratori Nazionali di Frascati, Italy}

\author{N. Sakaki}
\affiliation{RIKEN, Wako, Japan}

\author{K. Shinozaki}
\affiliation{Istituto Nazionale di Fisica Nucleare - Sezione di Torino, Italy}
\affiliation{Dipartimento di Fisica, Universit\'a  di Torino, Italy}
\affiliation{National Centre for Nuclear Research, Lodz, Poland}

\author{J. Szabelski}
\affiliation{National Centre for Nuclear Research, Lodz, Poland}

\author{Y. Takizawa}
\affiliation{RIKEN, Wako, Japan}

\author{S. Turriziani }
\affiliation{RIKEN, Wako, Japan}
\affiliation{Currently affiliated to  Gubkin Russian State University of Oil and Gas (NRU), Moscow, Russia}

\author{M. Tra\"{i}che}
\affiliation{Centre for Development of Advanced Technologies (CDTA), Algiers, Algeria}

\author{G. Valentini}
\affiliation{ASI, Italian Space Agency, Italy}

\author{S. Wada}
\affiliation{RIKEN, Wako, Japan}

\author{L.~Wiencke}
\affiliation{Colorado School of Mines, Golden, CO, USA}

\author{I. Yashin}
\affiliation{Skobeltsyn Institute of Nuclear Physics (SINP), M.V. Lomonosov Moscow State University, Moscow,
Russia}

\author{A. Zuccaro-Marchi}
\affiliation{RIKEN, Wako, Japan}
\affiliation{currently at ESA, Noordwijk, Netherlands}

\begin{abstract}
Mini-EUSO is a telescope observing the Earth in the ultraviolet   band from the International Space Station.
It is a part of the JEM-EUSO program, paving the way to future larger missions,
such as K-EUSO and POEMMA, devoted primarily  to the observation of Ultra High Energy Cosmic Rays  from space.
Mini-EUSO is capable  of observing Extensive
Air Showers generated by Ultra-High Energy Cosmic Rays with an energy above 10$^{21}$ eV and detect artificial showers generated with lasers from
the ground. Other main scientific objectives of the mission are the search for nuclearites and  Strange Quark Matter, the
study of atmospheric phenomena such as Transient Luminous Events, meteors
and meteoroids, the observation of sea bioluminescence and of artificial satellites and man-made space
debris.
Mini-EUSO will map the night-time Earth in the UV range (290 - 430 nm), with a spatial resolution of about 6.3 km and a temporal resolution of 2.5 $\mu$s, through a  nadir-facing UV-transparent window in the Russian Zvezda module.
 The instrument, launched on  August $22$, 2019 from the Baikonur cosmodrome, is based on an optical system employing two Fresnel lenses and  a focal surface composed  of 36 Multi-Anode Photomultiplier tubes, 64 channels each, for a total of 2304 channels with single photon counting sensitivity and an overall  field of view of 44$^{\circ}$.  Mini-EUSO also contains two ancillary cameras to complement  measurements in the near infrared and visible ranges. 
In this paper we describe the detector and present the various phenomena observed in the first months of operations.

\end{abstract}

\keywords{Mini-EUSO, JEM-EUSO, meteors, nuclearites, Strange Quark Matter, Space debris, bioluminescence, Transient Luminous Events, UHECR, Cosmic rays}

\section{Introduction} 
\label{sec:introduction}
After 80 years from the first measurement of  Extensive Air Showers (EAS) \citep{RevModPhys.11.288} and almost 60 years after the first report of a particle with energy of 10$^{20}$ eV \citep{PhysRevLett.10.146}, the origin and nature of  Ultra-High Energy Cosmic Rays (UHECRs), particles with E $\geq$ 10$^{19}$ eV, remain unsolved. This is mostly  due to the extremely low particle flux  – around 1 particle per km$^2$ per millennium - reaching the Earth at energies of the order of 10$^{20}$ eV. Currently, two ground-based observatories, the Pierre Auger Observatory (PAO) \citep{Aab_2020} and Telescope Array (TA) \citep{Abbasi_2018}, are observing the sky from the Southern and Northern hemisphere, respectively.  In the future,   an important step forward in studying UHECRs could come from space-based experiments which have the potential to look at the whole sky with a much larger equivalent active area \citep{Panasyuk_2015,ISI:000419565300021, 2013APh....44...76A}.
   
Observation of UHECRs from space   is based on the measurement of fluorescence and Cherenkov photons produced in EAS. A UHECR hitting  the atmosphere produces secondary particles that, in turn, collide with the air atoms producing a  shower  largely dominated by electrons and positrons. Crossing the atmosphere, these particles excite metastable energy levels in atmospheric molecules, especially nitrogen. When the electrons in these atoms return to the ground state, they  emit characteristic fluorescence light in the ultraviolet (UV) band, with wavelengths between 290 and 430 nm \citep{airfly}.
This light is emitted  isotropically, with an intensity  proportional to the energy deposited by the shower in the atmosphere.  The EAS thus forms a streak of fluorescence light along its path in the atmosphere, depending on the energy and zenith angle of the primary particle. Another detectable component  is the Cherenkov light emitted in the forward direction by the charged, relativistic  particles of the EAS and reflected into space by the ground or the clouds.  Looking   downward at the Earth's atmosphere from space, a specifically designed telescope can  detect the light emitted in the EAS  path.  At any atmospheric depth, the recorded amount of light is nearly proportional to the shower size at that point.  By imaging the motion of the UV track on timescales of microseconds or less, it is possible to define the arrival direction of the primary cosmic ray \citep{2015ExA....40..153A}. The integral of recorded light allows a determination of the energy of the primary UHECR. The shape of the shower, especially the position of the shower maximum in the traversed slant depth, gives a hint about the nature of the primary particle.

The most relevant advantage of space-based observations of UHECRs is the extremely large instantaneous observational area that can be monitored from space compared with that of the on-ground arrays. A second relevant feature of this approach is the  uniform exposure over the full sky. Space observations can cover a 4$\pi$ sky with the same instrument, ensuring identical experimental performance for southern and northern hemispheres. However, since they detect  fluorescence light and are thus operational in moonless nights, they have a low duty cycle ($\simeq 10\%$ \citep{Argiro:2003zb}) and have a larger systematic uncertainty compared to  ground based observatories which can also detect the charged component of the shower \citep{ABBASI2016131}.

\begin{figure}[ht]
\centering
\includegraphics[width=0.8\textwidth]{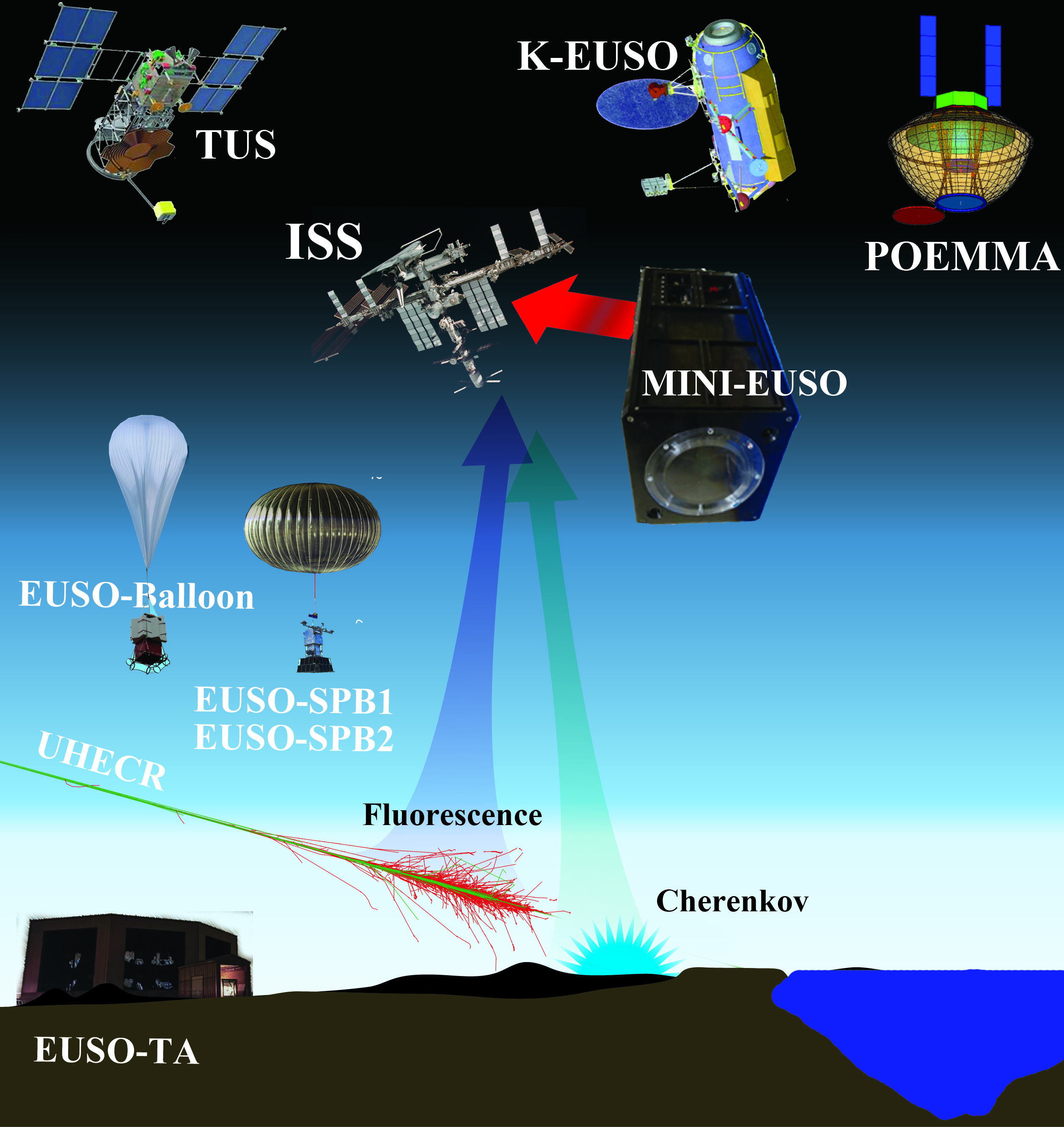}
\caption{Sketch of the JEM-EUSO Program.}
\label{fig-1}       
\end{figure}

The idea to go to space to study UHECRs by observing the fluorescence light produced by EAS in the Earth's atmosphere, was first proposed by John Linsley in 1981 \citep{1981ICRC....8..145B}. Since then, different attempts have been made to develop detectors to search for UHECRs from space \citep{1999ICRC....2..384S}. Currently, this issue is addressed by the JEM-EUSO (Joint Experiment Missions for Extreme Universe Space Observatory) program \citep{2017ICRC...35..370C}. This program, implemented by a collaboration of about  300 researchers from sixteen countries, includes several missions: EUSO-TA \citep{2018APh...102...98A} on ground, EUSO-Balloon \citep{2019APh...111...54A}, EUSO-SPB1 \citep{2017ICRC...35.1097W}  and  EUSO-SPB2 \citep{2017arXiv170304513A} on stratospheric balloons,  Mini-EUSO and TUS \citep{2017SSRv..212.1687K} in space, the medium size mission K-EUSO \citep{2017ICRC...35..368C} that will measure for the first time from space the UHECR flux at energies above 10$^{19}$ eV.   
In the roadmap of  future of   UHECR observation from space is the planned POEMMA \citep{2019BAAS...51g..99O}, a large NASA  project under evaluation. See Figure \ref{fig-1} for an overview of the various projects.
 
UHECR detectors in space  also allow the observation in the UV band of a number of phenomena ranging from the  detection of meteors and space debris, to Transient Luminous Events (TLEs) and emissions from terrestrial and marine surfaces.

\begin{figure}[ht]
\centering
\includegraphics[width=0.7\textwidth]{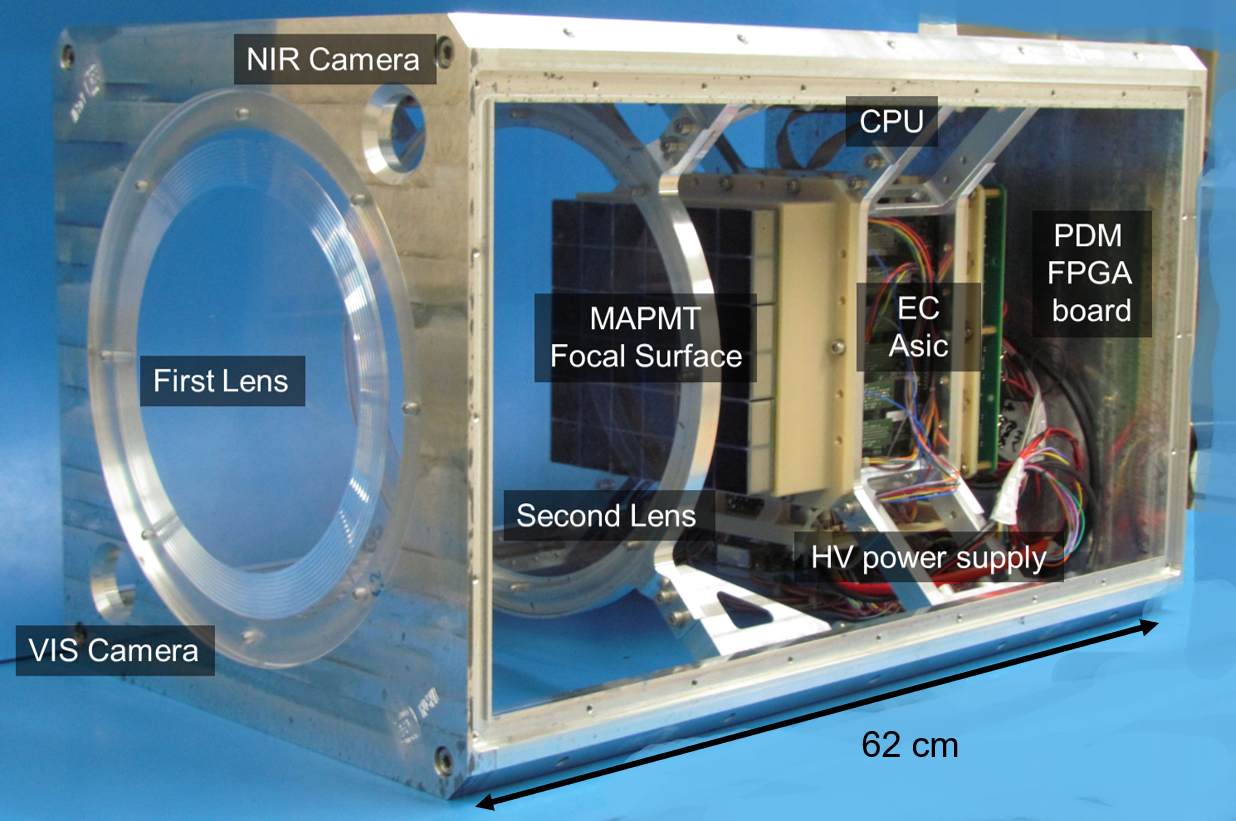}
\includegraphics[width=0.7\textwidth,angle=-90]{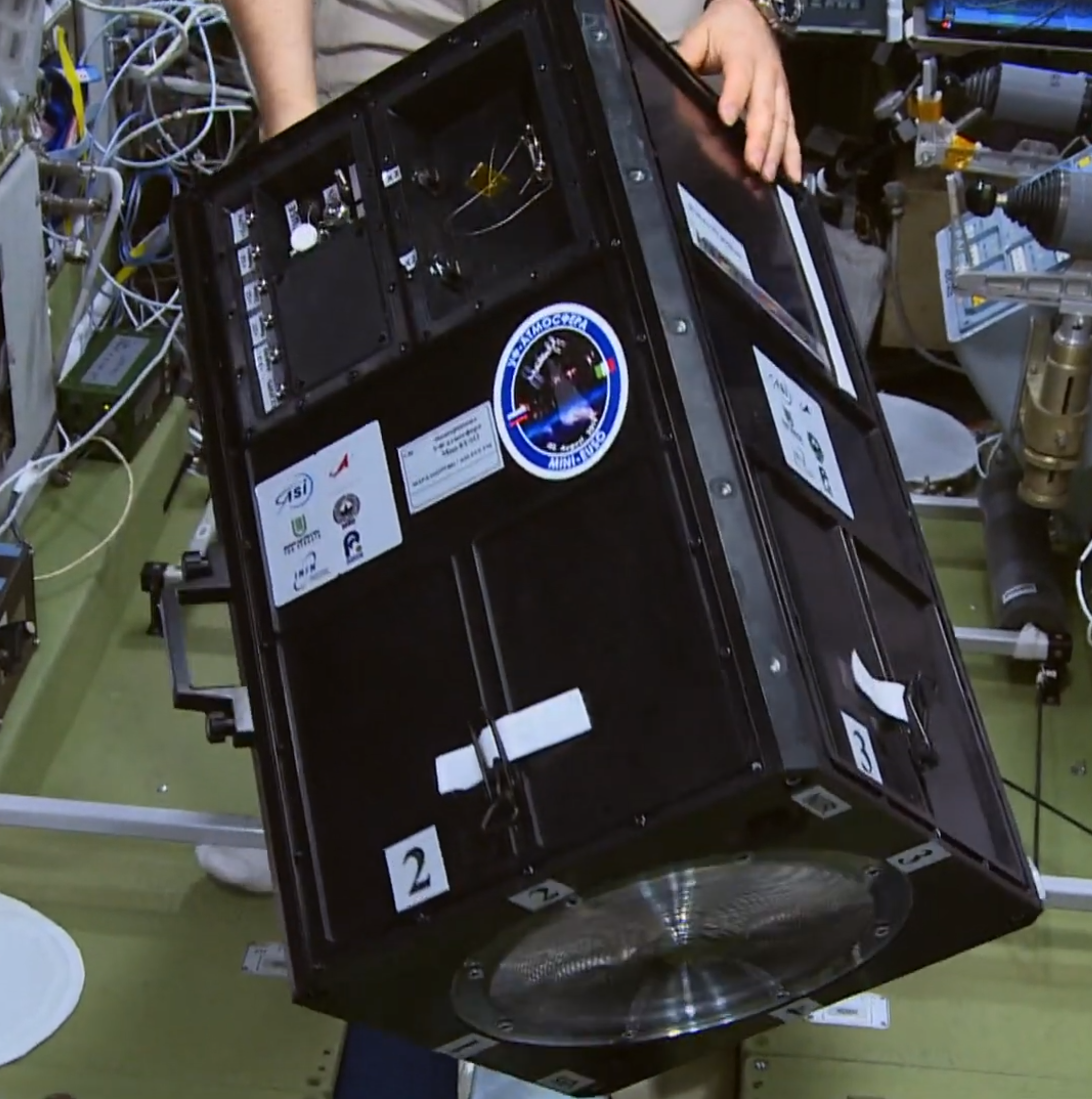}
\caption{Top: Mini-EUSO Engineering model (37x37x62 cm$^3$) during assembly, with the main elements visible \citep{Belov2020_AMeS}. Bottom: Mini-EUSO Flight model   in the ISS Zvezda module  prior to its installation on the nadir window. The panel on the top (marked as '2') houses the connectors for power supply, grounding and SSD mass storage. }
\label{fig-1a}       
\end{figure}

Mini-EUSO (Multiwavelength Imaging New Instrument for the Extreme Universe Space Observatory or  ``UV atmosphere'' in the Russian Space Program)    is a  telescope (Figure \ref{fig-1a})  operating in the UV range (290 - 430 nm) with a  square field of view of $\simeq $44$^{\circ}$ and a ground resolution of $\simeq 6.3 \times 6.3\:$ km$^2$  \citep{Capel:2018ig}, depending on the   altitude of the International Space Station (ISS). Mini-EUSO was brought to the  ISS  by the uncrewed Soyuz MS-14, on  August $22$, 2019. First observations from the  nadir-facing UV transparent window in the Russian Zvezda module  took place  on October $7$. Since then, it has been taking data periodically, with installations occurring every couple of weeks. The instrument is expected to operate for at least three years.  
The optical system consists of  two   Fresnel lenses with a  diameter of 25 cm. The focal surface, or Photon Detector Module (PDM), consists of 36 MultiAnode Photomultipliers (MAPMTs) tubes by Hamamatsu, 64 pixels each, capable of single photon detection. Readout is handled by ASICs (Application Specific Integrated Circuit) in frames of $2.5\: \mu s$ (this is defined as 1 Gate Time Unit, GTU). Data are then processed by a Zynq based FPGA board which implements a multi-level triggering, allowing the measurement of triggered UV transients for 128 frames at   time scales of both  $2.5\: \mu$s and $320\: \mu$s.  An  untriggered  acquisition mode  with $\simeq $ 40.96 ms frames performs continuous data taking.   

\section{Scientific objectives}\label{Sci_goals}

\begin{figure}[ht].
\centering
\includegraphics[width=0.8\textwidth]{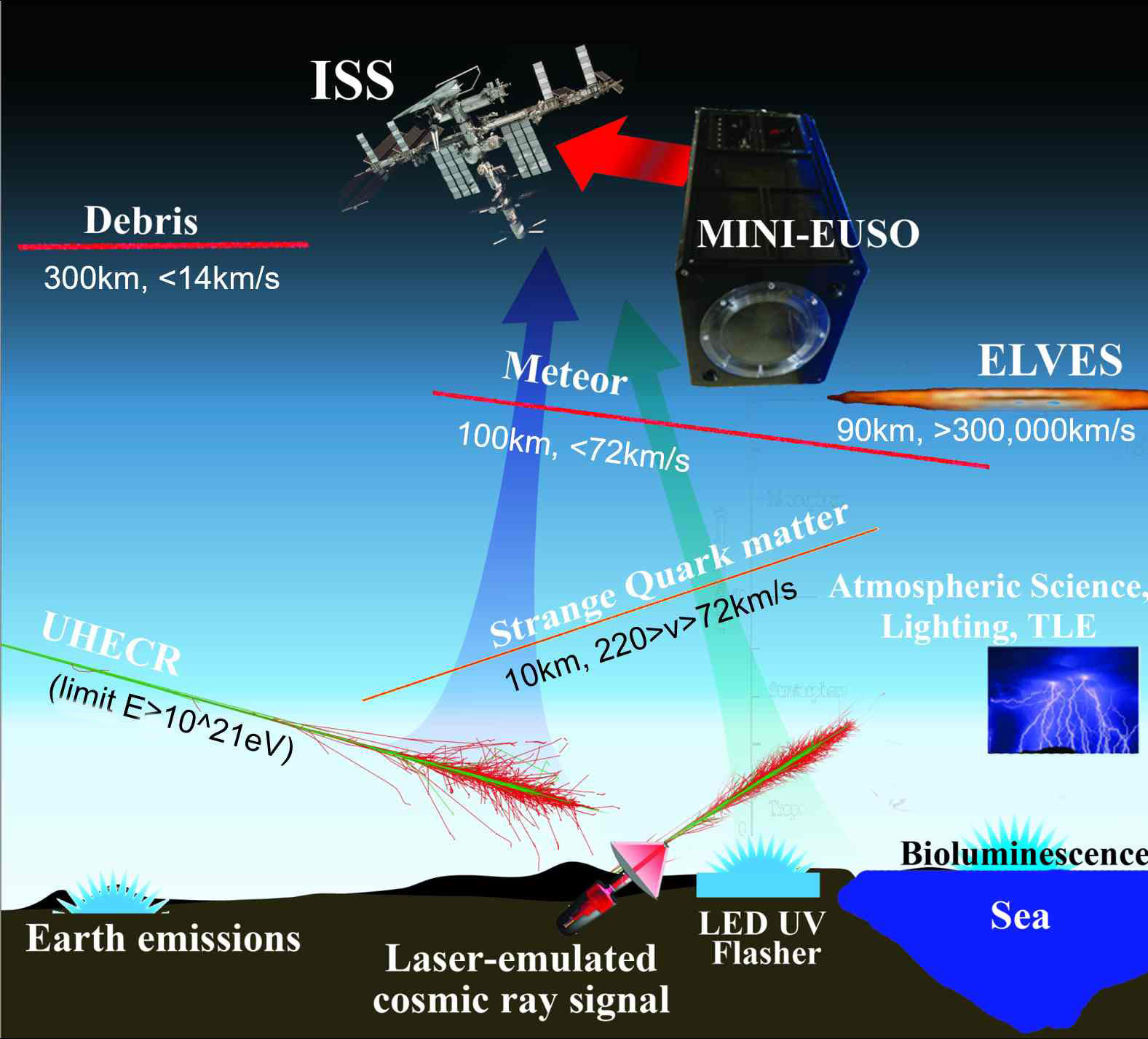}
\caption{Scientific objectives of the Mini-EUSO experiment. The detector is capable of addressing phenomena with greatly varying intensity and duration, from the slow terrestrial emissions to the apparently superluminal Elves (see text).}
\label{fig-scientificobjectives}       
\end{figure}

 Mini-EUSO is capable of addressing a number of different scientific objectives  occurring at various time scales. The main topics  - in order of decreasing duration of the phenomena -  are (Figure \ref{fig-scientificobjectives}):

\begin{enumerate}
\item{Night UV emissions from the Earth:} Mini-EUSO will map the Earth in the near  UV range   with   spatial and temporal resolution  of $\sim 6.3 \times 6.3 $ km$^2$ and 2.5 $\mu$s respectively, measuring variations of the UV emissions. This allows   studying the emissions from different surfaces, as snow, cloud, grass, savannah, etc..., taking into account also seasonal variations, which could be related to human activities.
These observations will lead to a better understanding of climatic effects and the presence of hot aerosols in the atmosphere, as well as monitoring volcano emissions. In particular, the observation of UV emissions over the seas and oceans, caused by algae or plankton bioluminescence \citep{Miller14181}, provides information on marine pollution. Similar studies over land give information on the status of terrestrial vegetation, measure the effects of human activities such as the potential of UV emissions by agricultural crops or the emissions generated by industrial or civilian facilities. 

\item{Airglow:}
in the chemistry of the atmosphere a particular role is played by airglows that -  in the wavelength band 330-400 nm that can be detected by Mini-EUSO -  are dominated by the emission from oxygen molecules in the Herzberg I band around the boundary region at an altitude of 95 km between the mesosphere and thermosphere \citep{2009JGRD..11420114K,2005AdSpR..35.1964I}.  Mini-EUSO will study the geographical and time variation of airglows, which are possible effects of geomagnetic disturbances in the upper atmosphere and also of  tsunami events. It will be possible, for example, to identify and reconstruct patterns of atmospheric gravity waves induced by tsunami waves in airglow light \citep{doi:10.1029/2002GL016069}. 

\item{Space debris:} attempts will be made to track space debris to investigate the possibility of using laser ablation for their removal. The maximum detection distance of Mini-EUSO is about 100 km for debris size of 0.1 m. This observation is restricted to the local twilight period of the orbit, about 5 min every 90 min. Since from the nadir-observing window the geometry of observation is not optimal, we are studying the possibility to change the observation window and inclination, as well as irradiate the detected debris using a prototype low power laser system. The successful irradiation will be confirmed by the brightening of the debris by the reflected laser photons \citep{Ebisuzaki2015102ActaAstronautica}.

\item{Meteors} are relatively slow ($v\leq 72$ km/s) and long-lasting  (a few seconds) events which  illuminate in sequence several light sensitive pixels of the Mini-EUSO focal surface.  Mini-EUSO will contribute to meteor hazard estimation by covering a projected surface on the ground of about 320 x 320 km$^2$ including inter-pixel dead areas (at 100 km altitude, where meteors burn, the field of view is about 240 x 240 km$^2$); it could be considered  a precursor for the optimization of future instruments for detecting meteors from space. The maximum observable magnitude is between 4 and 5 depending on background conditions \citep{2017P&SS..143..245AMeteorstudiesintheframeworkoftheJEMEUSOprogram,2015ExA....40..253A}. The time dependence of the light intensity will be determined for each event. The frequency-intensity distribution of the observation of meteors will allow to make an inventory of the population of near-Earth objects from space with a large field of view and the advantage of not being covered by clouds. In addition, when meteors are observed in coincidence with ground-based telescopes, it will be possible to determine their three-dimensional path and original heliocentric orbit  \citep{2015ExA....40..253A}.

\item{Strange Quark Matter} (SQM) is a theoretically predicted bound state of up, down, and strange quarks \citep{1984PhRvD..30..272W}. Thanks to the lowering of the Fermi energy per nucleon by the addition of a third quark species,  this matter could be  stable \citep{1993PhRvL..71..332G} and form the true ground state of hadronic matter.  SQM could have cosmological origin and might be present at the core of neutron stars \citep{2014PhRvD..89d3014D, 2007Natur.445E...7A}.  In particular, strange stars, made of SQM and bound by the strong interaction, should exist with very different properties from those predicted for hadronic neutron stars.  If  SQM nuggets - also called nuclearites - are present in our Galaxy they could encounter the Earth and interact with its atmosphere.   The high density of these nuclearites would produce a long and constant signal in the atmosphere \citep{DeRujula:1984axn}. Furthermore,  their interstellar origin would result in higher speed than solar system meteors. These two features  should permit a clear identification of this class of events \citep{Piotrowski:2020ftp}. As mentioned, Mini-EUSO is able to observe meteors down to magnitude 4 and  5, corresponding to a SQM nugget size of $10^{24} $ GeV/c or higher\citep{2015ExA....40..253A}.

\item{Transient Luminous Events} (TLEs: sprites, blue and gigantic jets, halo, Elves etc.) are upper atmospheric optical phenomena of electromagnetic nature, connected with thunderstorms \citep{2006seil.book.....F}.   Elves (Emission of Light and Very low frequency perturbations due to Electromagnetic pulse Sources) were  predicted \citep{doi:10.1029/91GL00364}  before observation \citep{doi:10.1029/91GL03168, Fukunishi1996}, and last a few hundred $\mu s$ at most.  The global coverage of its measurements allows Mini-EUSO to observe these events and study the so-called far-from-thunderstorm transient atmospheric events, observed previously in a couple of experiments \citep{8358705, doi:10.1029/2019EA000582}.
  Furthermore, it will be possible to make joint observations with other detectors on board the ISS such as Altea-Lidal \citep{Rizzo_2019}  and ASIM \citep{2019SSRv..215...23O}. ASIM has observed several Elves and  reported the emissions of a Terrestrial Gamma-ray Flash (TGF) in conjunction with an Elve \citep{Neubert183}.
 Using the Altea-Lidal data  will make it possible to correlate particle counts with signals observed by Mini-EUSO in order to study correlations.

\item{UHECR:} Mini-EUSO can measure fluorescence and Cherenkov light emitted by UHECR initiated showers.   The  diameter of the lens system - constrained by the size of the ISS window - places the threshold energy for UHECR detection  around  10$^{21}$ eV.  According to  Auger \citep{PhysRevLett.101.061101, Abraham:2010ks}, TA \citep{Abu-Zayyad:2013gf} and joint Auger-TA \citep{Deligny:2019tb, 10.1093/ptep/ptx054} data, it is therefore unlikely to observe these events. However, in order to pave the way to future larger systems and optimize their design, a UV ground flasher and a UV ground laser will be operated during fly-by \citep{ADAMS20141506, Adams:2016UL}. Moreover, Mini-EUSO should shed light on the nature of the extreme energy EAS-like event recently reported by TUS \citep{Khrenov:2017eev}, by observing several of these events thanks to its larger field of view \citep{Khrenov_2020}. 

\item{Exposure for future UHECR and neutrino observations from space:} Mini-EUSO will provide a large variety of essential data to estimate the exposure of space-based detectors for the detection of UHECRs  and neutrinos from space, namely the intensity of UV terrestrial emissions and its variation as a function of time and geographic location; the role of anthropogenic lights, the lightning frequency, their contribution to the increase of the UV light intensity as well as the possibility of localizing them in UV. The effect of these factors was originally taken into account in the calculation of the exposure for JEM-EUSO \citep{2013APh....44...76A, UHECR3, UHECR5}. Mini-EUSO data will provide fundamental information to evaluate to evaluate the exposure more accurately.

\end{enumerate}

\section{Instrument Overview}\label{InstView}
Mini-EUSO has been designed to be installed in the interior of the ISS on the UV-transparent window located in the Zvezda module. The dimensions (37x37x62 cm$^3$) are thus defined by the size of the window and the constraints of the  Soyuz spacecraft. Furthermore, the design accommodates the requirements of safety (no sharp edges, low surface temperature, robustness...) to the crew.   Coupling to the window is done   via a mechanical adapter flange; the only connection to the ISS is via a 28 V power supply and grounding cable. The power consumption of the telescope is $\simeq60\:$ W and the weight is 35 kg, including the  5 kg flange.  
For each observation  session, taking place  about every two weeks and of the duration of about 12 hours,  the instrument is removed from storage and installed on the  UV window. Data are stored   on 512 GB USB Solid State Disks (SSD)  that are inserted in the side of the telescope by the astronauts.  No direct  telecommunication with ground is present, but each session samples of data (about 10$\%$, usually corresponding to the beginning and the end of each session) are copied by the crew and transmitted  to ground to verify the correct functioning of the instrument and optimize its working parameters. Conversely, before each session, working parameters, patches in software and hardware are uplinked to the ISS and then copied on the SSD disk to fine-tune the acquisition of the telescope. Pouches with 25 SSDs are   returned    to Earth every 6 months.

The telescope can be divided in  3 main systems: the optics, the focal surface and the data acquisition.

\subsection{Optical System}
\begin{figure}[ht]
\centering
\includegraphics[width=0.8\textwidth]{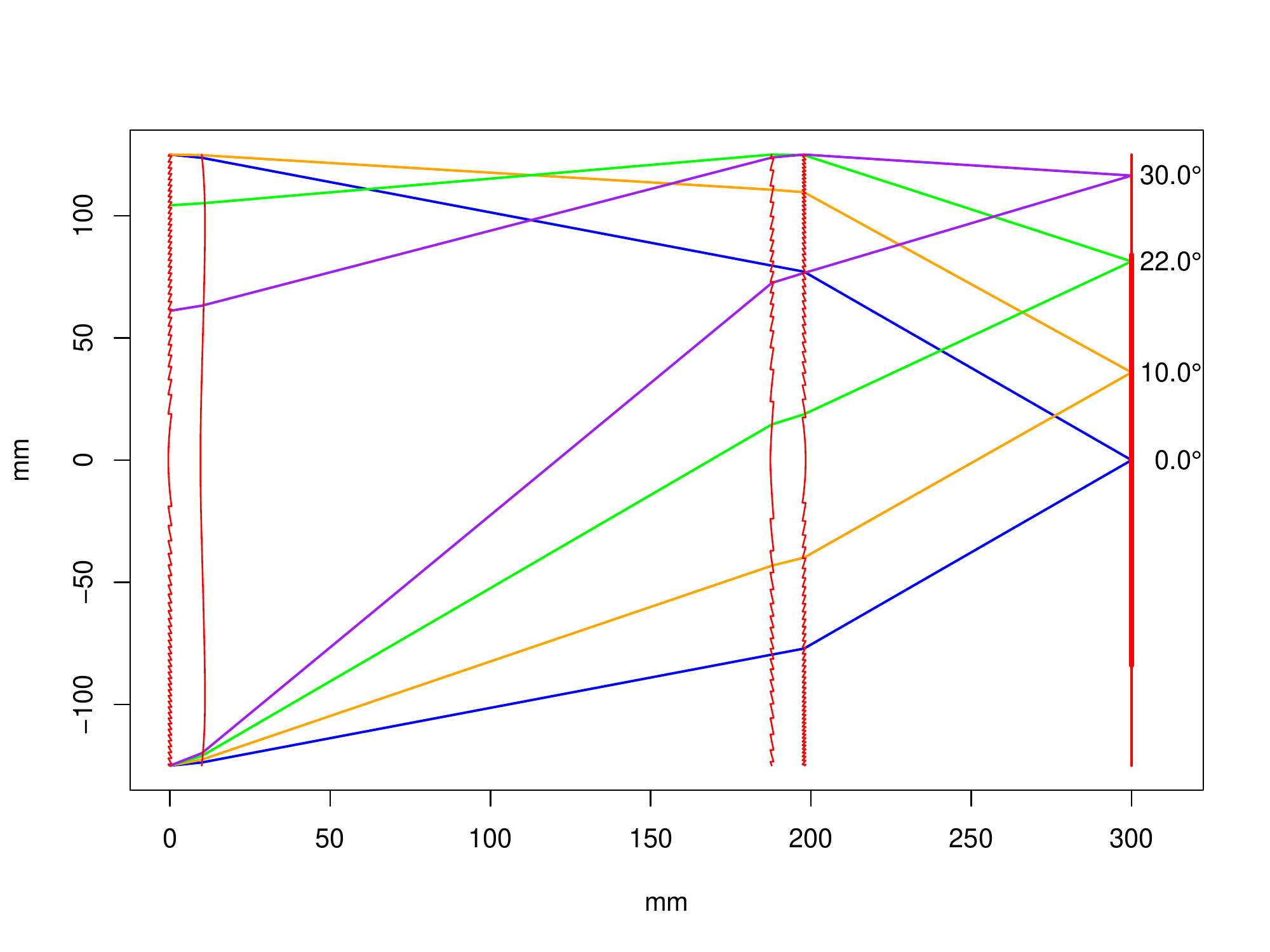}
\caption{Mini-EUSO optical system configuration. Light enters from left of the picture, crosses two,  11 mm thick, lenses with a focal length of 300 mm and reaches the focal surface (PDM) to the right.}
\label{fig-6}       
\end{figure}

The optics (Figure \ref{fig-6}) consists of two, 25 cm diameter, Fresnel lenses with a wide field of view (44$^{\circ}$ seen from the PDM).   Poly(methyl methacrylate) - PMMA - is used to manufacture the lenses with a diamond bit machine. In this way it is possible to have a  light (11 mm thickness, 0.87 kg/lens), robust and compact design well suited for space applications.  The effective focal length of the system is 300 mm, with a Point Spread Function (PSF) of 1.2 pixels, of the same dimension as the pixel size of the MAPMTs.  

\begin{figure}[ht]
\centering
\includegraphics[width=0.8\textwidth]{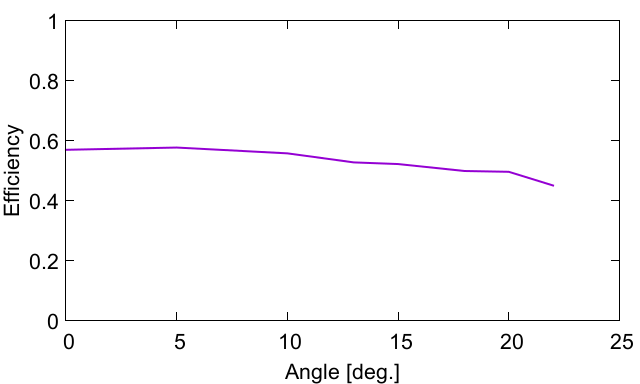}
\caption{The photon collection Efficiency (PCE) of the Mini-EUSO lens system. The PCE has been measured with a 405 nm laser  as a function of the angle at which photons enter the first lens, and is defined as the light collected  in  a square of  5 mm size. 
}
\label{fig-PCE}        
\end{figure}

The measured Photon Collection Efficiency (PCE) of the optical system, defined as the number of photons which arrives in one pixel  divided by the number of photons incident upon the front lens, is plotted  in Figure \ref{fig-PCE}. The figure shows   the good transmittance and uniformity of response of the system, exhibiting  some vignetting effect only at large  angles, corresponding to the borders and corners of the focal surface.

\subsection{The Focal surface }

\begin{figure}[ht]
\centering 
\includegraphics[width=0.8\textwidth]{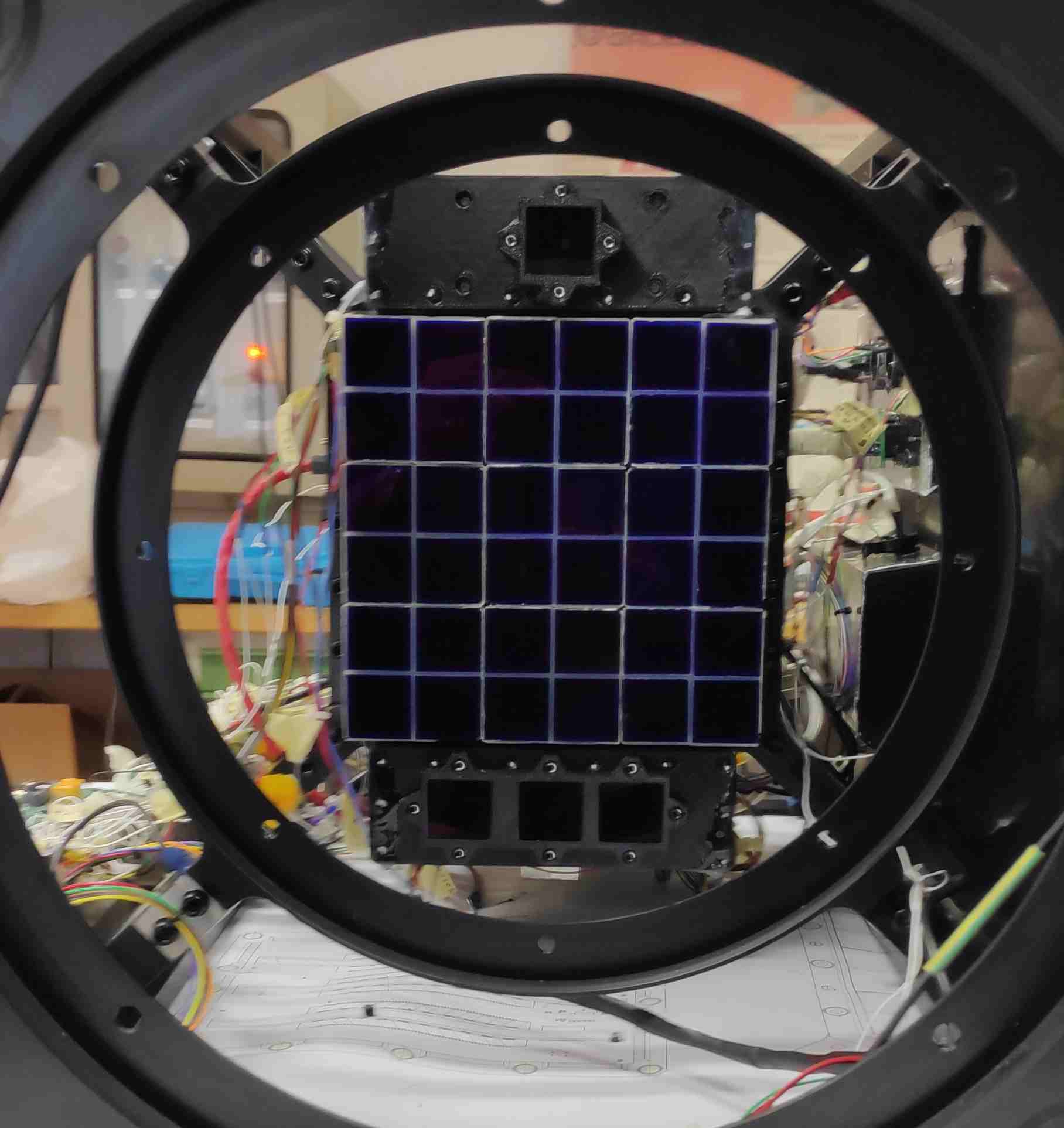}
 \caption{Mini-EUSO focal surface. The  Photo Detector Module (PDM) is composed by 36 Multi-Anode Photomultiplier Tubes, each with 64 independent channels (2304 total pixels) and arranged in groups of four (an Elementary Cell, EC). On top of the PDM is a 64 channel Silicon Photomultiplier, at the bottom of the PDM are two light sensors and a single-pixel SiPM.}
\label{fig_FS}      
\end{figure}

\begin{figure}[ht]
\centering
\includegraphics[width=0.3\textwidth]{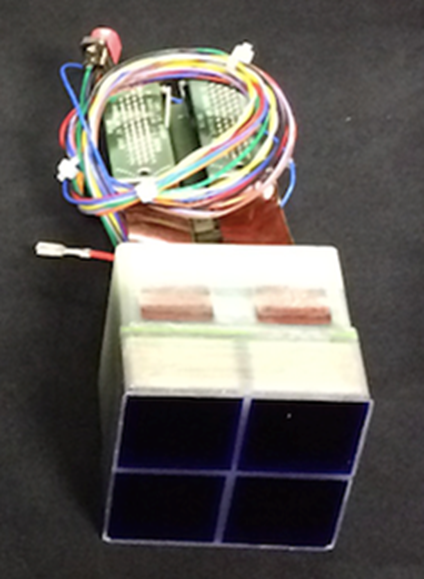}
\caption{Picture of an Elementary Cell (EC) unit. It consists of four MAPMTs (Hamamatsu Photonics, model R11265-M64), each covered with a 2 mm BG3 UV filter with anti-reflective coating. }
\label{ECfig}        
\end{figure}

\begin{figure}[ht]
\centering
\includegraphics[width=0.8\textwidth]{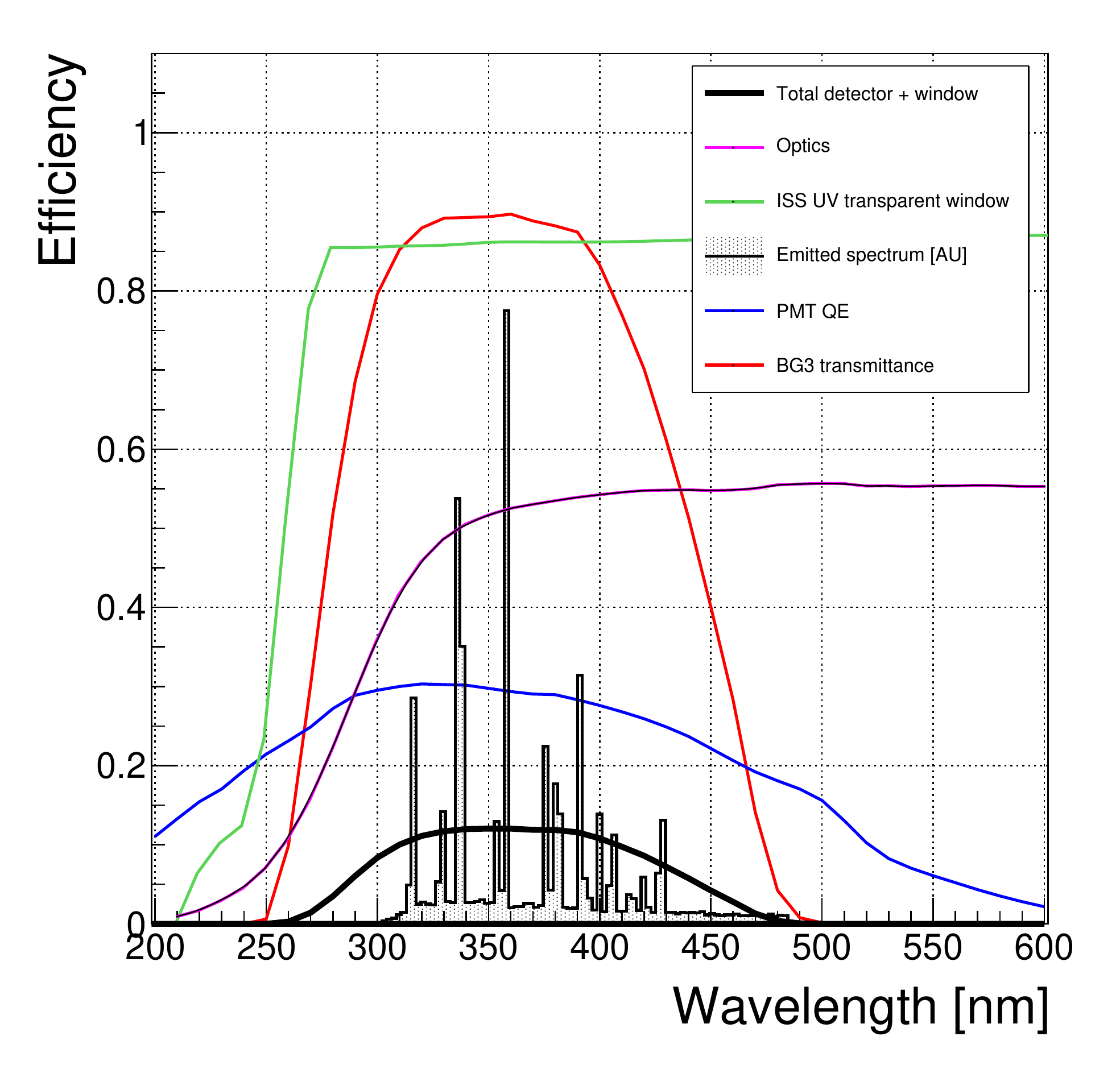}
\caption{Efficiency of the Mini-EUSO detector (black curve) as a function of wavelength. The total detector efficiency (black line)  is the result of the Optics transmittance (purple curve), the transmission of the BG3 bandpass filter (red curve), the PMT Detector Efficiency  (DE) of the photocathodes (blue curve) and of the UV transparent window of the ISS (green curve). The system has been designed to optimize observations of the fluorescence light emitted by nitrogen atoms excited by the extensive air shower of cosmic rays (grey histogram).}
\label{efficiency}         
\end{figure}

The Mini-EUSO focal surface (PDM) consists of a matrix of 36 Multi-Anode Photomultiplier Tubes (MAPMTs, Hamamatsu Photonics  R11265-M64), arranged in an array of 6$\times $6 elements. Each MAPMT consists of 8$\times$ 8 pixels, resulting in a total of 2304 channels (Figure \ref{fig_FS}).
The MAPMTs are grouped in Elementary Cells (ECs), each with $2\times2$ units. Each of the nine  ECs  (Figure \ref{ECfig}) of the PDM shares a common high voltage power supply and a board connecting the dynodes and anodes of the four photomultipliers. The whole system   (250 g each EC, including filters and MAPMTs) is  potted with Arathane and located in the shadow of the photosensors. 
 Similar PDM units have been used in the ground telescope of EUSO-TA \citep{2018APh...102...98A} and in the first two balloon flights, EUSO-Balloon \citep{2015ExA....40..281A,2019APh...111...54A}  and  EUSO-SPB1 \citep{2017ICRC...35..384B}. A more complex setup, involving three PDMs side-by-side, will be used in the upcoming EUSO-SPB2 flight \citep{2017arXiv170304513A}. 

UV bandpass filters (2 mm of BG3 material) with anti-reflective coating are glued in front of the MAPMTs  to select mostly wavelengths between 290 nm and 430 nm. In Figure \ref{efficiency} are shown the various contributions to the overall detector efficiency. The Detection Efficiency (DE) of the MAPMTs has been obtained rescaling the Quantum Efficiency curve provided by Hamamatsu by a typical collection
efficiency of 80$\%$. The result is consistent with the detection efficiency measured in laboratory at 398 nm.

The array of MAPMTs in the Mini-EUSO PDM is powered by a low-power consumption Cockroft-Walton high voltage power supply (HVPS). The system has an internal safety circuit which removes the electric potential difference between the photocathode and the first  dynode in case of high current drain due to bright light (more than 100 counts in a given GTU on more than three pixels of a given EC). If the high current drain  persists due to very bright sources (e.g. strong lightning), the  HV for that EC unit is turned off. 

In this way, the system allows  Mini-EUSO to  observe phenomena with brightness varying   several orders of magnitude:  from a UHECR showers (a few counts/pixel/GTU) to bright Elves ($\simeq $ 80 counts/pixel/GTU) and lightning (up to and even beyond 10$^4$counts/pixel/GTU). 



\subsection{The Data Acquisition Module}
  Spaciroc-3 ASICs \citep{2018NIMPA.912..363B} are used as front-end electronics. Each Spaciroc-3 handles in parallel 64 independent channels and thus preamplifies and digitizes photoelectron signals  from a single MAPMT. 
	  The MAPMTs are operated in photon counting mode  to minimize the contribution of the integrated noise. When a UV photon hits the photo-cathode, a photo-electron, p.e., is produced with a probability depending on the quantum efficiency of the cathode. After being amplified with a gain of the order of 10$^6$ in the dynode cascade in the photomultiplier, the signals are discriminated (with a threshold for each PMT) and integrated in a  2.5 $\mu s$ period (defined - as mentioned - to be 1 GTU). Single photon discrimination is 5 ns.  
Every GTU each  ASIC sends the counts from the corresponding MAPMT to the   PDM Data Processing (PDM-DP) system for readout. The PDM-DP is  based on a  Zynq board containing a Xilinx FPGA and an embedded dual core ARM9 CPU processing system. The board is responsible for the majority of the data handling, from data reception,   buffering, configuration of the Spaciroc-3 ASICs, implementation of the  trigger algorithms and interfacing with the  CPU. The HVPS board is also controlled by this module in order to have a real-time response to high light signals as a second safety level against bright light. 
The PDM-DP stores the 2.5 $\mu s$ data stream (D1) in a running buffer on which runs the trigger code. The algorithm  searches for a signal above 16 standard deviations from the average in any pixel of the focal surface. Both the rms and the average are calculated in real time to take into account varying illumination conditions. In case of a trigger, the 128 frame buffer (64 frames before the trigger and 64 after it) is stored in memory. Independently from the trigger, sums of  128 frames (320 $\mu s$, D2) are continuously calculated and  temporarily stored in another buffer where  a similar  trigger algorithm (at this time scale) is run. See \citep{2018AdSpR..62.2966B} for a more detailed description of the trigger  logic. Similarly, sums on 128 D2 frames (40.96 ms, D3) are also stored in real time. Every $ 5.24$ s 128 packets of D3 data and up to  4 D2 packets and 4 D1 packets (if triggered) are sent to the CPU for storage.

Key parameters of the trigger algorithm, such as the threshold and integration period, are configurable and can be changed in-flight by modifying the contents of the SSD. 

\subsection{CPU} 
 
\begin{figure}[ht]
\centering
\includegraphics[width=0.8\textwidth]{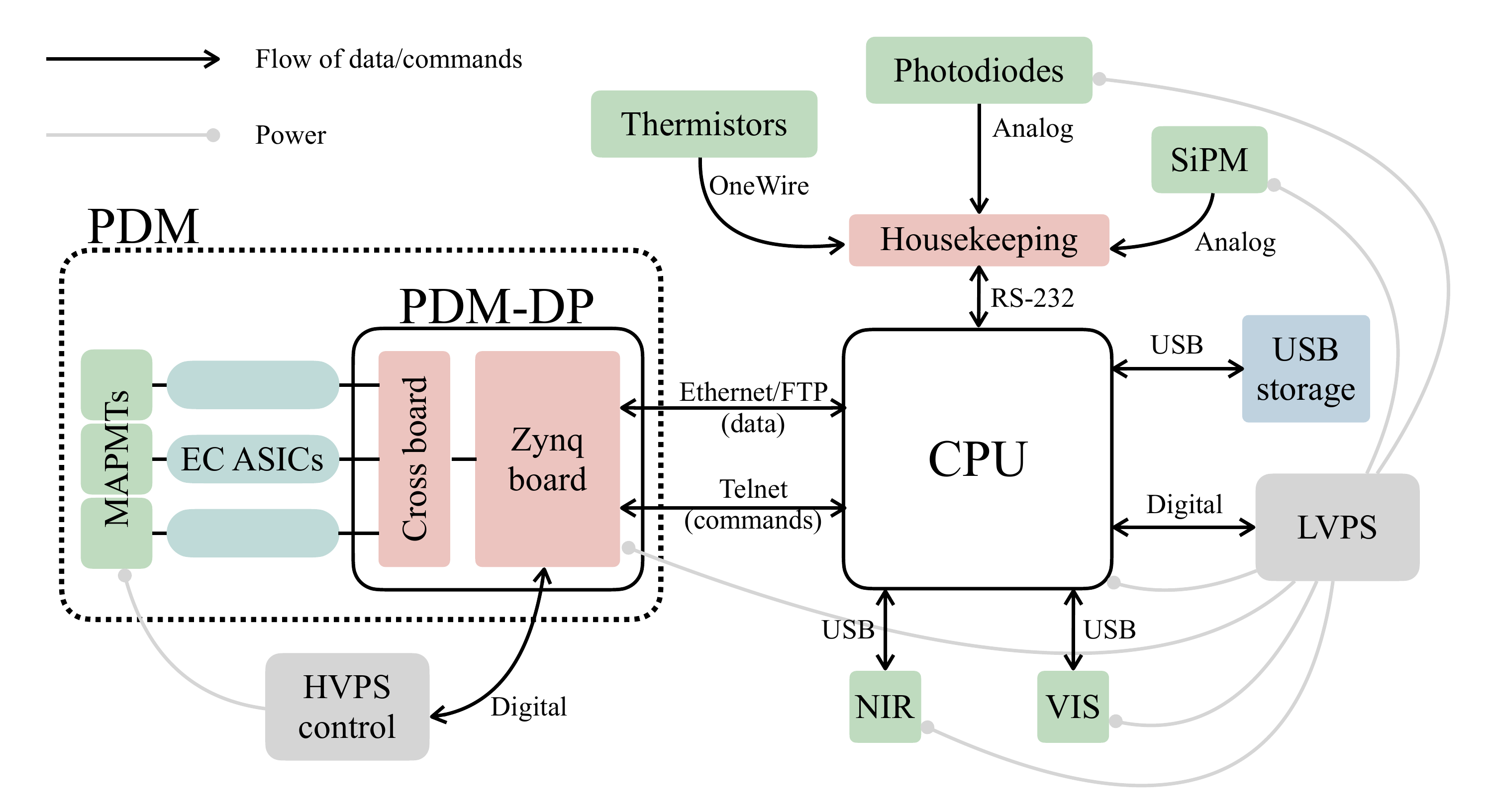}
\caption{Mini-EUSO Data Acquisition System scheme with the main interfaces. Incoming data from the Spaciroc-3 ASICs is triggered in the Zynq board
and then passed to the CPU via an ethernet link. The CPU combines this data with that of the housekeeping system and the ancillary cameras. Data is then moved to onboard SSDs for storage.}
\label{fig-CPU}        
\end{figure}

The CPU (CMX34BT, 1.33 GHz single core Atom) is devoted to the task of controlling the instrument, handling the data management and storage on SSD cards,  housekeeping, switching between operational modes and collecting data from the NIR (Near InfraRed)  and  VIS (Visible) cameras (in 4-second frames) as well as the ancillary systems \citep{ISI:000510649500019}. 
The data acquisition system is summarized in Figure \ref{fig-CPU}.

\subsection{Ancillary systems}
Mini-EUSO houses two  cameras, one in the near infrared (NIR, 1500 - 1600 nm) and one in the visible (VIS, 400 - 780 nm) band, to provide additional information in different frequency ranges. These cameras are located in the corners of the plane facing the window and are read directly by the CPU (USB bus). The data is acquired independently of the PDM \citep{2019AdSpR..64.1188T} in 4-second exposure frames . 
 Mini-EUSO sensors also include  a 64 channels Multi-Pixel Photon Counter (MPPC) Silicon PhotoMultiplier (SiPM) (Hamamatsu C14047-3050EA08) array, a single pixel SiPM (Hamamatsu C13365) and  two UV sensors (Analog Devices AD8304ARUZ, Lapis Semiconductor ML8511) for day/night information. These detectors are located in the focal plane of the PDM (see Figure \ref{fig_FS}) and read by the CPU.

\section{Engineering and flight models}\label{Models_and_QT} 
Two copies of the   detector have been realized: the Engineering Model (EM) and the Flight Model (FM).  In the EM only the four central MAPTMs are present and the other tubes are replaced with mass dummies. Both  copies  have been subjected to a series of qualification and acceptance tests, more severe for the EM.
The EM was used in the vibration,  Electric and Electromagnetic Interference and Compatibility (EMI-EMC), thermal-vacuum/environmental tests. The EM is now being used as a training model for the various  crews who will operate it on the ISS. 
The  FM underwent electric and vibration (at a reduced level) tests. See \citep{Belov2020_AMeS} for a detailed description of the qualification tests.

\subsection{Pre-flight tests }\label{Mech_QT}
The mass of the telescope required it to be launched  in a hard-mounted configuration inside the Soyuz capsule. This translates in higher vibration loads on the hardware. Shock and random vibration launch load tests, reflecting the corresponding requirements,  took place  in February and May 2019. The EM was subjected to random vibration tests  along the three axes in the frequency range from 20 to 2000 Hz, and shock acceleration tests  up to 40 g along  the three axes.
After each vibration sequence, visual inspection and pre/post resonance comparison to search for displacement of natural frequencies due to internal damage or loosening of parts was carried out.

EMI-EMC tests are performed to  verify that the Mini-EUSO instrument does not produce any undesired electromagnetic radiated emissions and that, conversely, it is capable of withstanding external electromagnetic interference. The EM  has thus been subjected to emission and susceptibility tests in  an anechoic chamber to verify its electromagnetic compatibility requirements: Low and High Frequency, conductive interference, electrical field intensity produced by high frequency emissions, pulse interference and inrush current. 

 Temperature and humidity conditions which might be experienced during the cargo transportation to the launch site and  within the Soyuz capsule have been reproduced in a thermal-vacuum chamber. 
During transportation to the Baikonur launch site, temperature excursions can range between $\pm$50 $^{\circ}$C, depending on the time of the year,  while the humidity can reach a level up to 90\%.  In the Soyuz, atmospheric conditions (450 to 970 mm Hg) are maintained during launch. 
Several thermal cycles 
inside a thermal chamber were  made at low and high temperatures, $\pm$55 $^{\circ}$C, and with humidity levels up to 95\%. 
After  each test  the instrument was switched on for a functional run.

\subsection{Tests on trigger}

 Trigger tests were conducted in 2018 at the TurLab facility \citep{bib:turlab} of the Department of 
Physics of the University of Turin, which hosts a $\sim$5 m diameter rotating tank used to perform 
analysis with moving light sources. 
The  facility  is located in a   dark environment 
where the intensity of background light can be adjusted and controlled. The Mini-EUSO EM, with a classical plano-convex lens
of 2.5 cm diameter, was hung on the ceiling above the TurLab tank and  tested there.

\begin{figure}[h]
	\vspace{-0.2cm}
	\centering
	\includegraphics[width=0.8\textwidth]{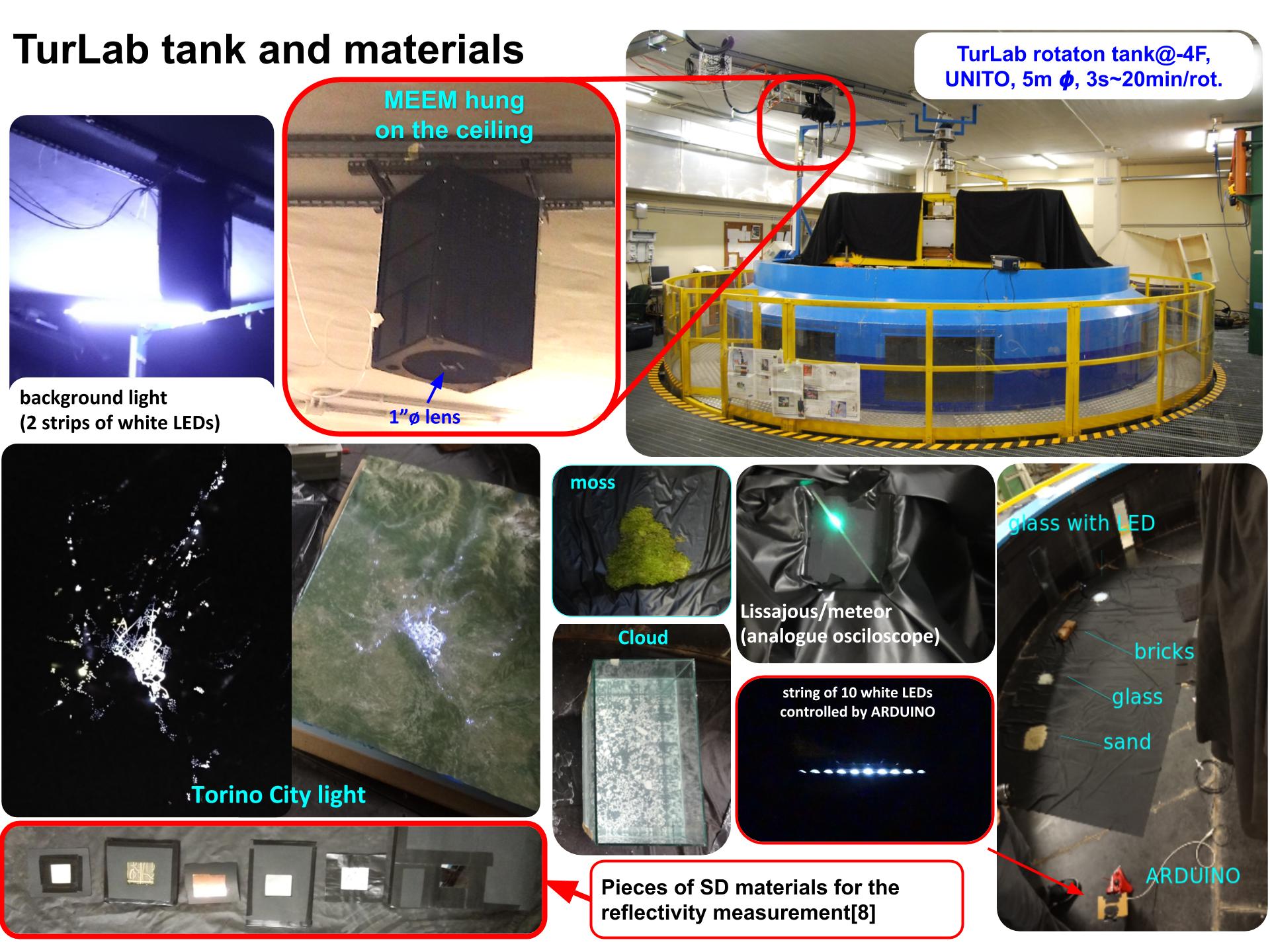}
	\caption{TurLab tank (top-right) setup with Mini-EUSO EM (top-centre) hung on the ceiling, light sources and materials to reproduce phenomena that the  Mini-EUSO detector 
can observe in space. Reflections of the materials which often become space debris are also tested (bottom-left). 
\label{fig:TurLabMat}
	}
	\vspace{-0.2cm}
\end{figure}

Figure~\ref{fig:TurLabMat} shows the setup for the TurLab measurement. Other pictures 
in the figure show the light sources and materials reproducing the various phenomena 
that Mini-EUSO can observe from space, such as rocks, desert, glacier ice, cloud, 
forest, lightning, city light as well as meteors and cosmic-rays. Each phenomenon is 
respectively reproduced   by means of bricks, sand, smashed glass, clusters of particles floating in 
the water, moss, LED light through the holes of a model of the city of Turin, 
Lissajous tracks on oscilloscope  and an Arduino driven LED strip.
All these materials were illuminated by diffused background light which was 
placed on  the ceiling above the tank, to reproduce the diffused air glow in the 
atmosphere at the level of the expected photon counts of $\sim$1~count/pixel/GTU. 
A Mini-EUSO observation along  the ISS orbit is reproduced by the rotation of the tank with the various light emitting or reflecting materials.

Additional outdoor observations were performed from the roof of the
 Department of Physics (Lat. $45^{\circ}03'08''$N, Long. $7^{\circ}40'53''$E) and at the 
Astronomical Observatory of Pino Torinese (Lat. $45^{\circ}02'25''$N, Long. $7^{\circ}45'53''$E), where the sky conditions allowed observation of faint 
sources. During these tests, flashers, building lights, stars with apparent magnitude up to 4 and Jupiter 
 have been observed. 
Four possible meteors have been detected, and from the comparison with the brightness of stars they have apparent magnitudes of about  4. 

Artificial lights from the urban area were also used to test the detector. 
Lighted signs and flashes of skyscrapers and towers located in the Turin area, as well
as airplane flashes were detected with D1 and D2 data streams. Light-curves with time 
sampling  of 320~$\mu$s show an alternating signal due to the 
$\sim$50~Hz signal form the electrical grid of the town.   
Light-curves with time bins of 40.96~ms show high pulses due to flashers of 
skyscrapers pulsating with a frequency of $\sim$0.7 Hz. 

\begin{figure}[h]
	\centering	
	\includegraphics[width=0.3\textwidth]{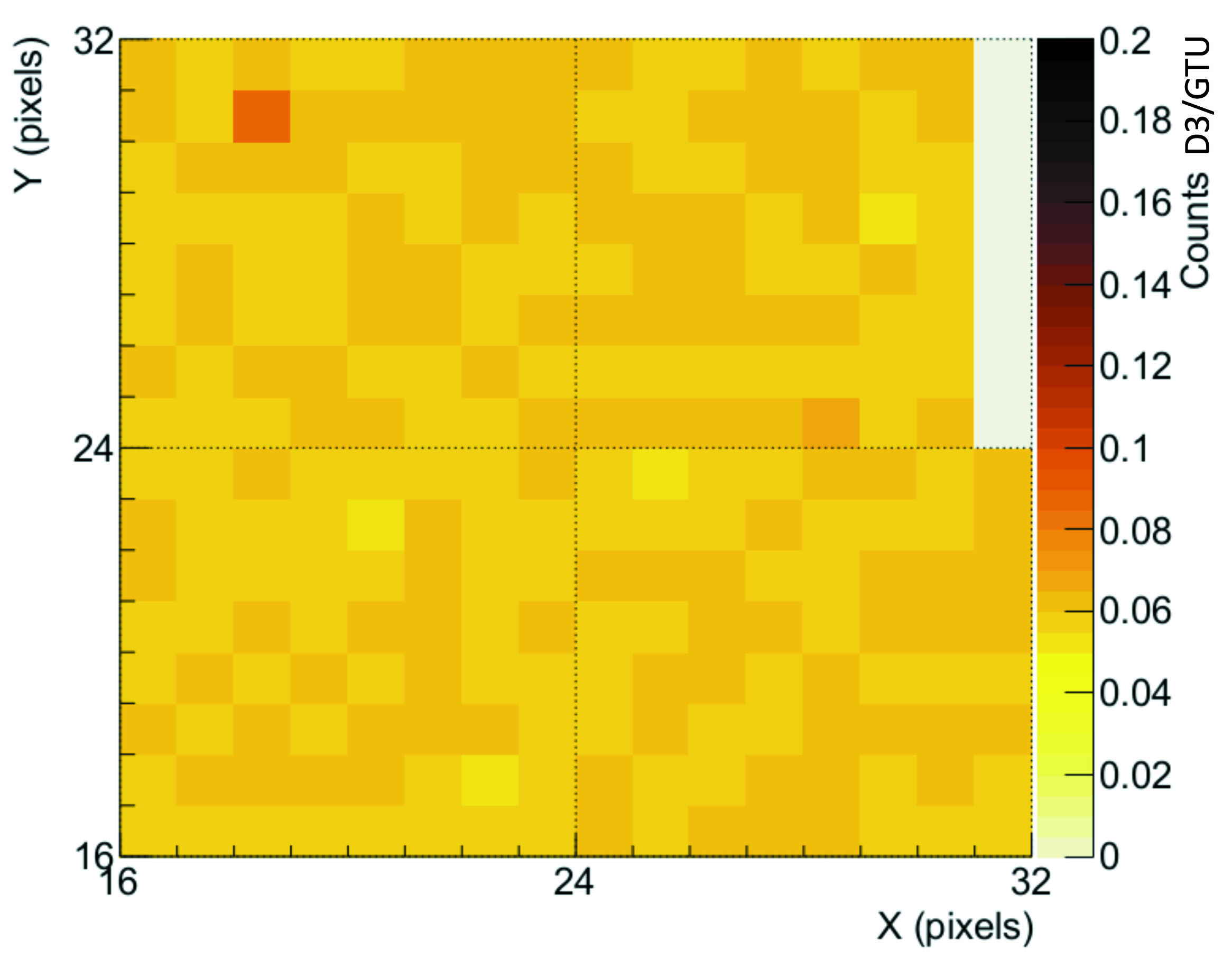}
	\includegraphics[width=0.3\textwidth]{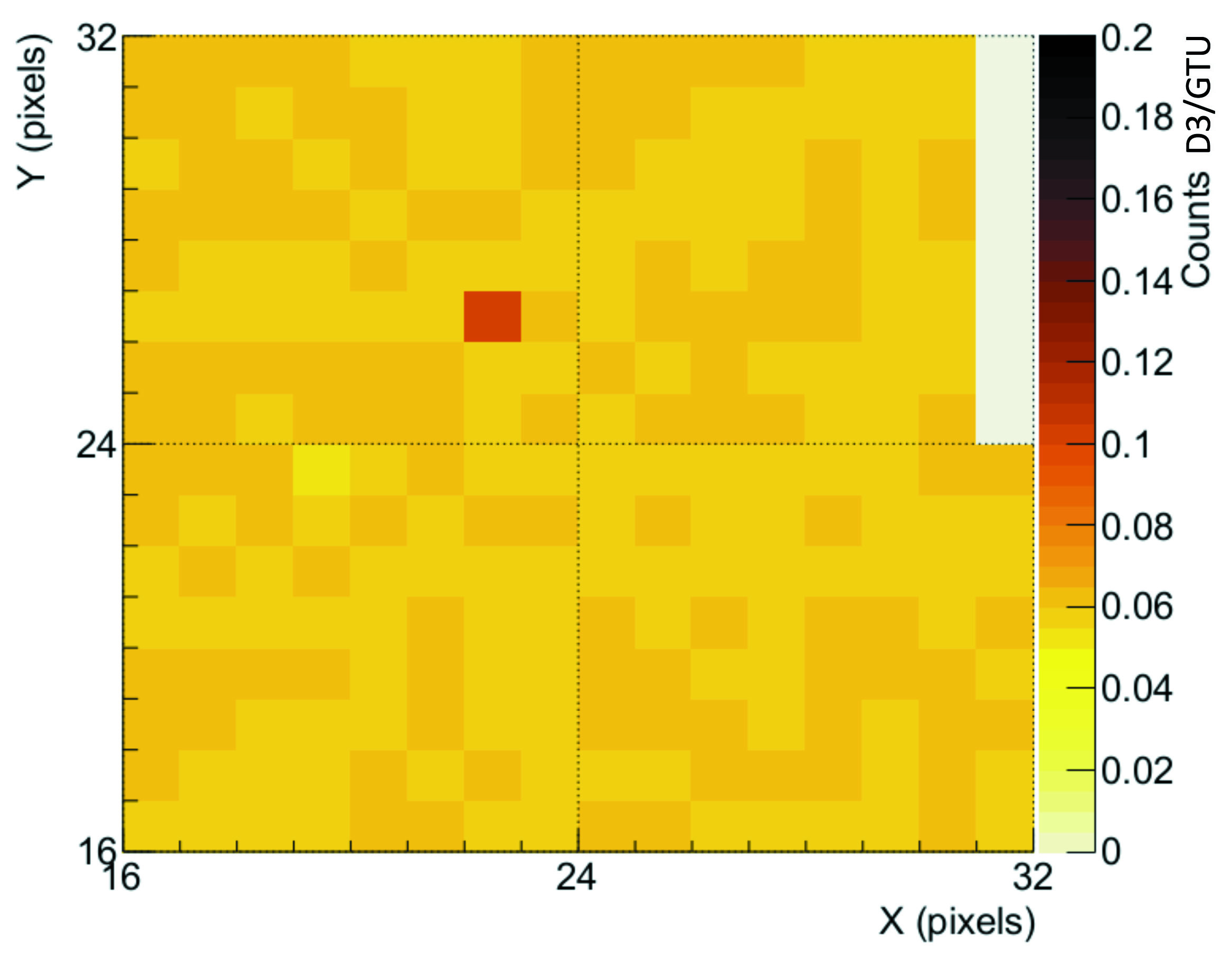}
	\includegraphics[width=0.3\textwidth]{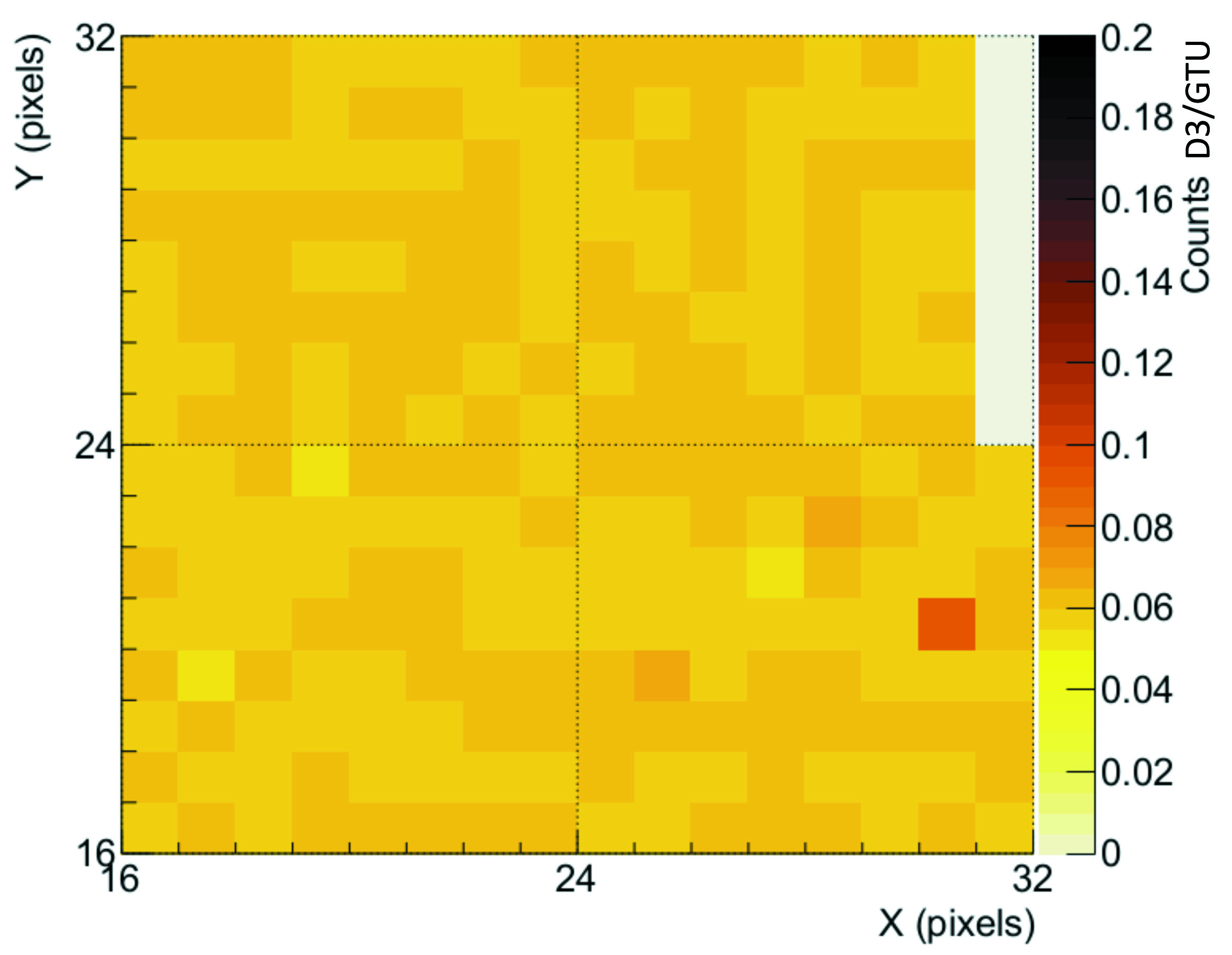}
	\caption{``Meteor 1-31 Rocket'' appearing in D3 frames (40.96~ms). Colour denotes counts/GTU (2.5 $\mu$s).} \label{fig:satellite}
\end{figure}

An orbiting rocket body that transported a telecommunication satellite was 
also detected and  later identified as the ``Meteor 1-31 Rocket''. It is a 2.6~m$\times$2.8~m 
large-size space debris orbiting at an average height of $\sim$550~km and an angular 
velocity of 0.78 $^\circ$/s. Figure~\ref{fig:satellite} shows the satellite as a bright 
pixel moving from the top-left corner towards the bottom-right corner of the frame in three
data frames integrated over 40.96 ms. 

\subsection{Field Tests}

 The instrument was field-tested in the Apennine  Mountains, close to the town of Paganico Sabino (Lat.  $42^{\circ}09'38 ''$N,  Long. $13^{\circ}00'33''$E), and in Rome from the roof of the Physics Department of the University of Rome Tor Vergata (Lat. $41^{\circ}51'15''$N, Long. $12^{\circ}36'15''$E).
		
\begin{figure}[h!]
\centering
\includegraphics[width=0.8\textwidth]{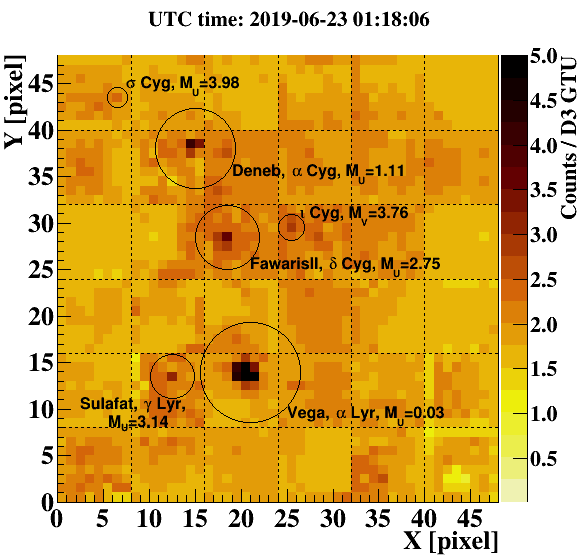} 
\caption{Mini-EUSO night acquisition frame acquired at a field test on June 23, 2019 (D3 acquisition). The brightest visible star is Vega ($M_u$=0.03),  the dimmest identified star is $\sigma$ Cyg ($M_u$=3.98).  Colour denotes counts/GTU (2.5 $\mu$s).}
\label{fig-12}
\end{figure}

In Figure \ref{fig-12} a night sky frame acquired in zenith position in the Apennine Mountains during ground tests is shown.
From the observation of the various stars in the field of view it is possible to obtain a first estimation of the PSF of the instrument of about 1.2 pixels, in agreement with the theoretical estimate (Figure \ref{fig-6}).

\subsection{Acceptance and Launch}

Finally, the detector passed several acceptance tests, first in Rome, subsequently in Moscow and finally in Baikonur cosmodrome. It was then integrated in the uncrewed  Soyuz capsule and launched on August    22, 2019. The first docking attempt on the Pirs module docking port, on August 24, 2019, was unsuccessful. After relocation of the Soyuz MS-13 capsule, a second, successful,  docking attempt on the Zvezda docking port took place on  August 28,  2019.

\section{In-flight Operations}\label{flight}

\begin{figure}[ht]
\centering 
\includegraphics[width=0.8\textwidth]{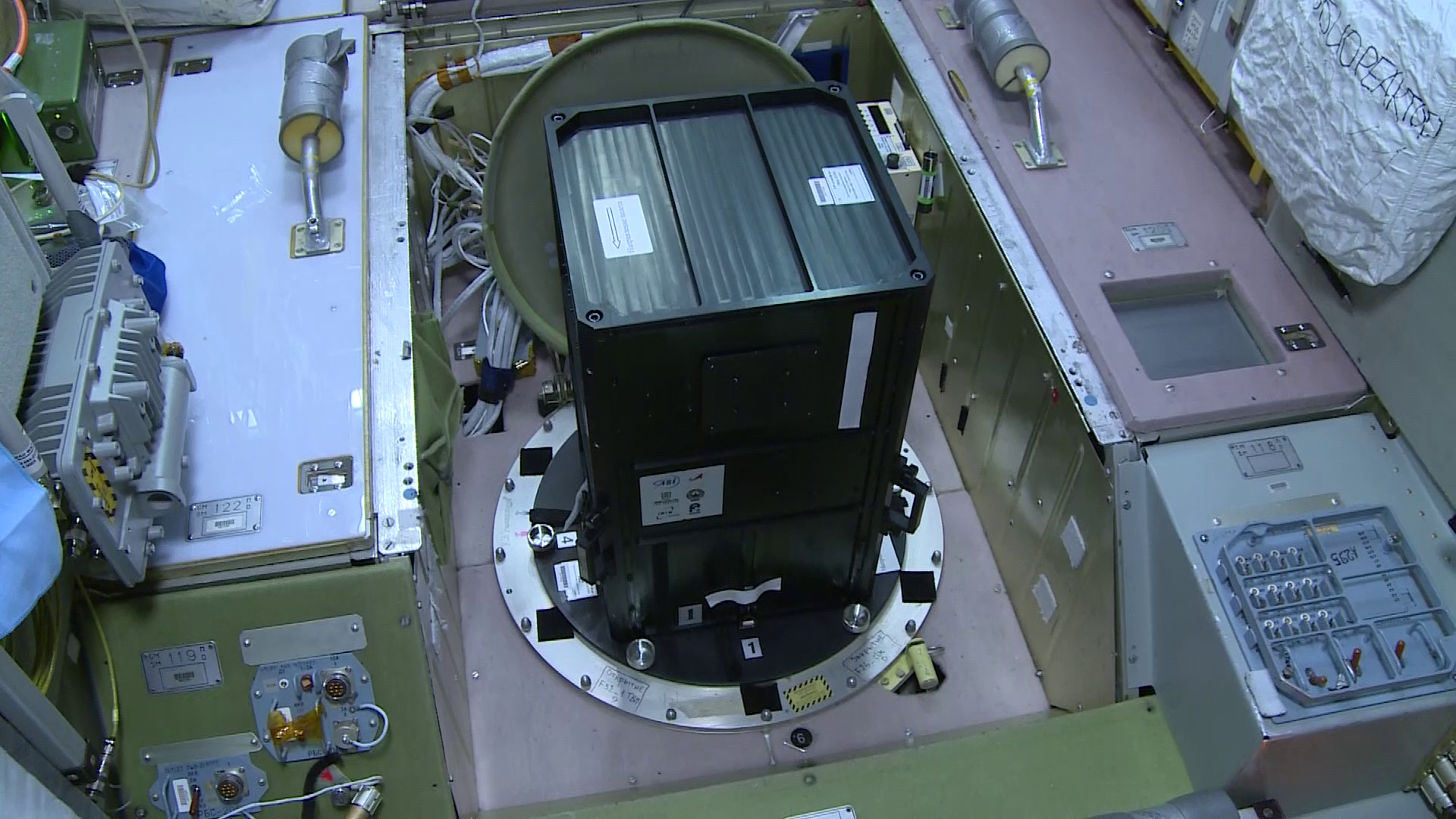}
 \caption{Mini-EUSO mounted on the UV transparent window of the Zvezda module. The velocity vector is usually toward the bottom of the picture, in the side marked as '1'.  }
\label{issops}       
\end{figure}

\begin{figure}[ht]
\centering 
\includegraphics[width=0.8\textwidth]{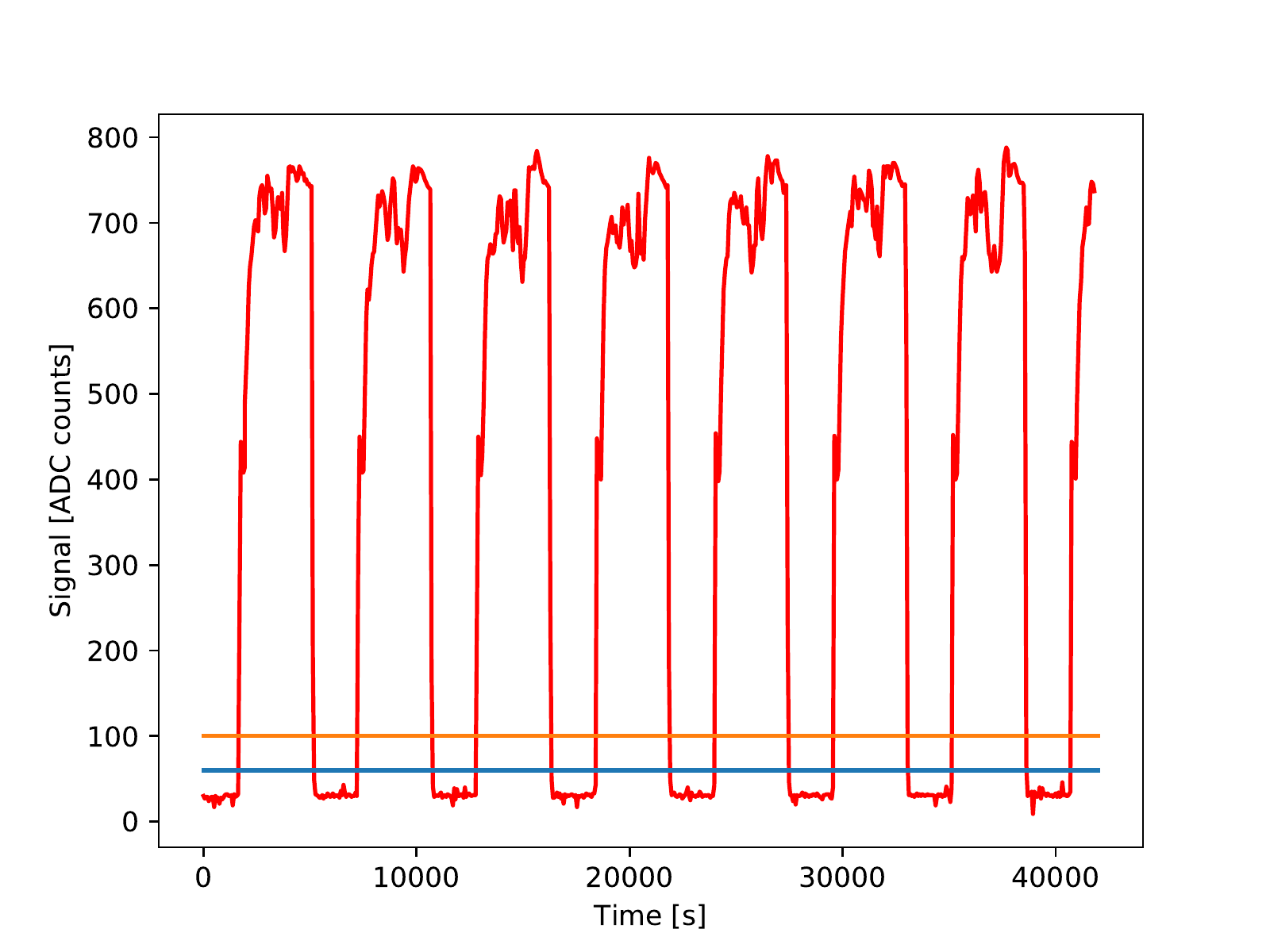}
 \caption{Plot of the measurements of the UV sensor as a function of time. Mini-EUSO is operational during night-time, when the sensor measures a value below 60 ADC (Analog-to-Digital Conversion)  counts. To avoid fluctuations at the day-night terminator line, two thresholds are used to determine transition from day to night (60 ADC counts, blue line) and vice-versa (100 ADC counts, orange line). }
\label{uvsensor}       
\end{figure}

The telescope was first turned on  October 7, 2019 (Figure \ref{issops}). As already mentioned, the detector is designed to operate in night-time conditions. The CPU handles cycling between day and night based on the measurements performed by the UV sensors located in the focal surface. The ML8511 sensor is  used for this purpose, although for redundancy reasons  all three sensors (two photodiodes and one SiPM) can be used.  To avoid spurious fluctuations between the two states at the night/day terminators,  two thresholds are used. Figure \ref{uvsensor} shows the light measured by the UV sensor as a function of time during one session of data taking. It is possible to see the transition between day and night every $\simeq$45 minutes\footnote{The illumination period depends on the Beta angle of the ISS, the angle between the orbital plane of the station and the Sun-Earth vector. When $\beta$ is close to $90^o$ (for ISS $\beta_{max}=75^o$) the station is almost always illuminated by the Sun and operations are not possible. When $\beta=0^o$ the duration of the local night is the longest.}. 

At the start of each session, the detector is taken from storage, the lens cover removed and the instrument is placed in position on the UV transparent window in the Zvezda module. Power and ground cables are connected, an SSD card is installed and power is switched on. Time is kept internally with a Real Time Clock  as there are no other connections with the ISS. The daily drift of the clock has been measured on ground and is periodically checked with data taken on board. 
Upon startup, the system checks whether  specific operational parameters which override the existing ones  are present on the SSD card. The initialisation program also checks if  software and/or firmware upgrades are present in the SSD and in that case it uses them. This flexible approach allows for continuous improvement of operations.

At the end of each session the detector is stored and the log file and a few data files are transmitted by telemetry for analysis and verification of the correct functioning of the system. 

The first session involved operation in safe mode, with only one EC unit active and the HVPS set  to last dynode voltage mode, corresponding to a sensitivity of about 1$\%$ compared to the normal HVPS mode. Gradually, along the course of the following sessions, the subsequent acquisitions have used the full PDM in normal voltage mode. 

\section{First observations}

Raw data are processed and analyzed with the ROOT \citep{BRUN199781} framework. 

\begin{figure}[ht]
\centering
\includegraphics[width=0.8\textwidth]{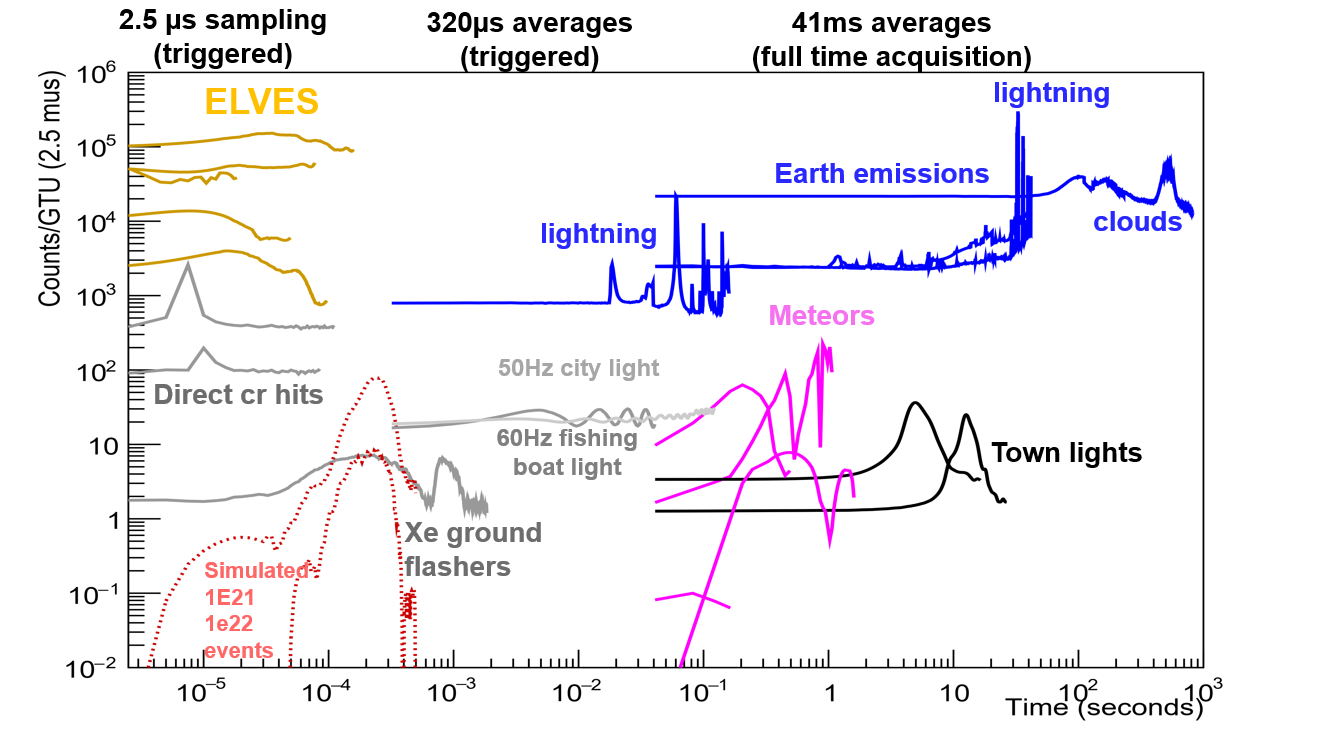}
\caption{Temporal profile of various signals observed by Mini-EUSO. }
\label{overview}        
\end{figure}

Figure \ref{overview} shows the observed total signal of the focal surface as a function of time for signals of various time scales, from the faster $2.5 \mu$s sampling (D1) to the 128 frame average for D2 (320 $\mu s$) to the 128$\times$128 frame average for D3 (40.96 ms).  In the longer time frames, the gradual increase is due to the passage over a clouded area, whereas the sharp spikes are due to lightning. Large lightning triggers the safety  system of the detector, resulting in the temporary deactivation of the HVPS of the EC unit which is overexposed by lightning.

\begin{figure}[ht]
\centering
\includegraphics[width=0.8\textwidth]{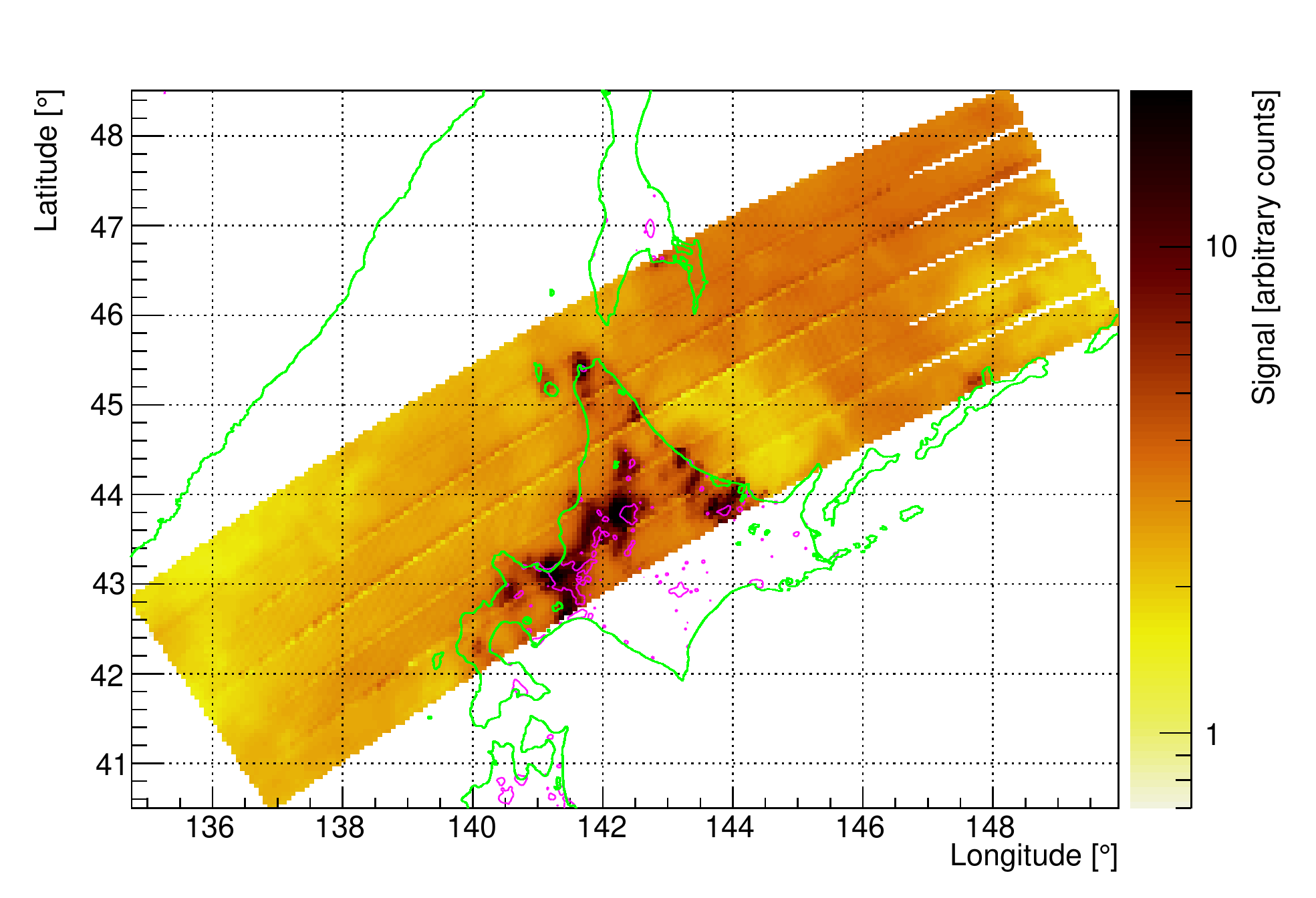}
\includegraphics[width=0.8\textwidth]{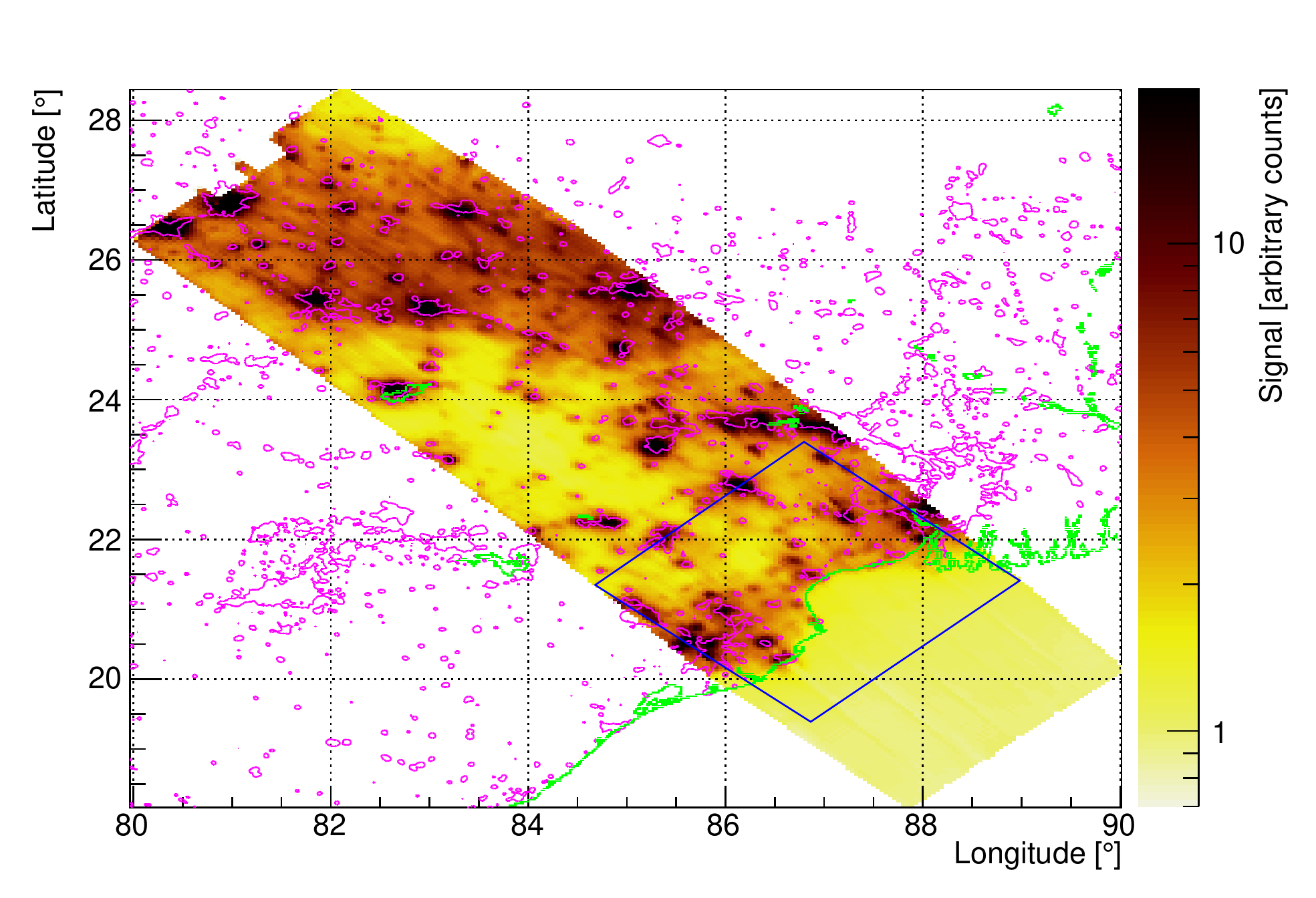}
\caption{Night time emissions of the Earth. Top:  Hokkaido, Bottom:  East coast of India. The green lines show the ground-water boundary, the purple lines show the emissions in the visible range (arbitrary units) \citep{dmsp}.  The square region shown in the coast of India shows the field of view of one D3 frame of Mini-EUSO and corresponds to the region shown in Figure \ref{frame}.}
\label{fig-nighttime}       
\end{figure}

\begin{figure}[ht]
\centering
\includegraphics[width=0.3\textwidth]{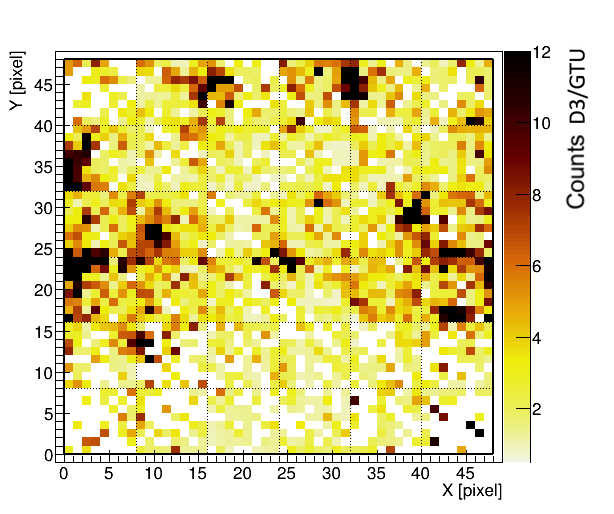}
\includegraphics[width=0.3\textwidth]{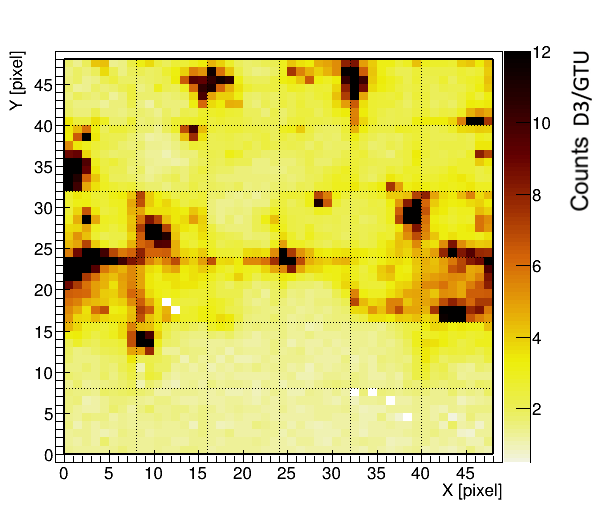}
\includegraphics[width=0.3\textwidth]{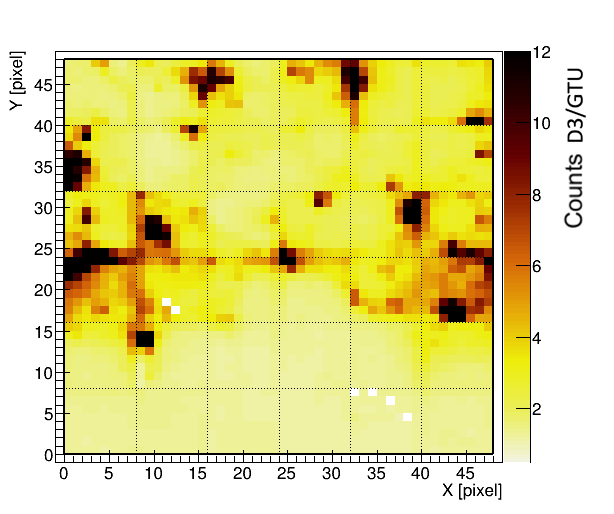}
\caption{Left: Frame taken with a 2.5 $\mu$s (D1) time sampling. Center: Frame taken with a 320 $\mu$s (D2) time sampling. Right: Frame taken with a 40.96 ms (D3) time sampling.  In all frames colour denotes counts/GTU (2.5 $\mu$s). The statistical fluctuations of D1  are averaged in D2 (128 frames) and D3 (128$\times$128 frames) images so that they appear to differ very little.  The sequence refers to approximately the same time of acquisition from the passage over India shown in Figure \ref{fig-nighttime}.}
\label{frame}       
\end{figure}

\begin{figure}[ht].
\centering
\includegraphics[width=0.7\textwidth]{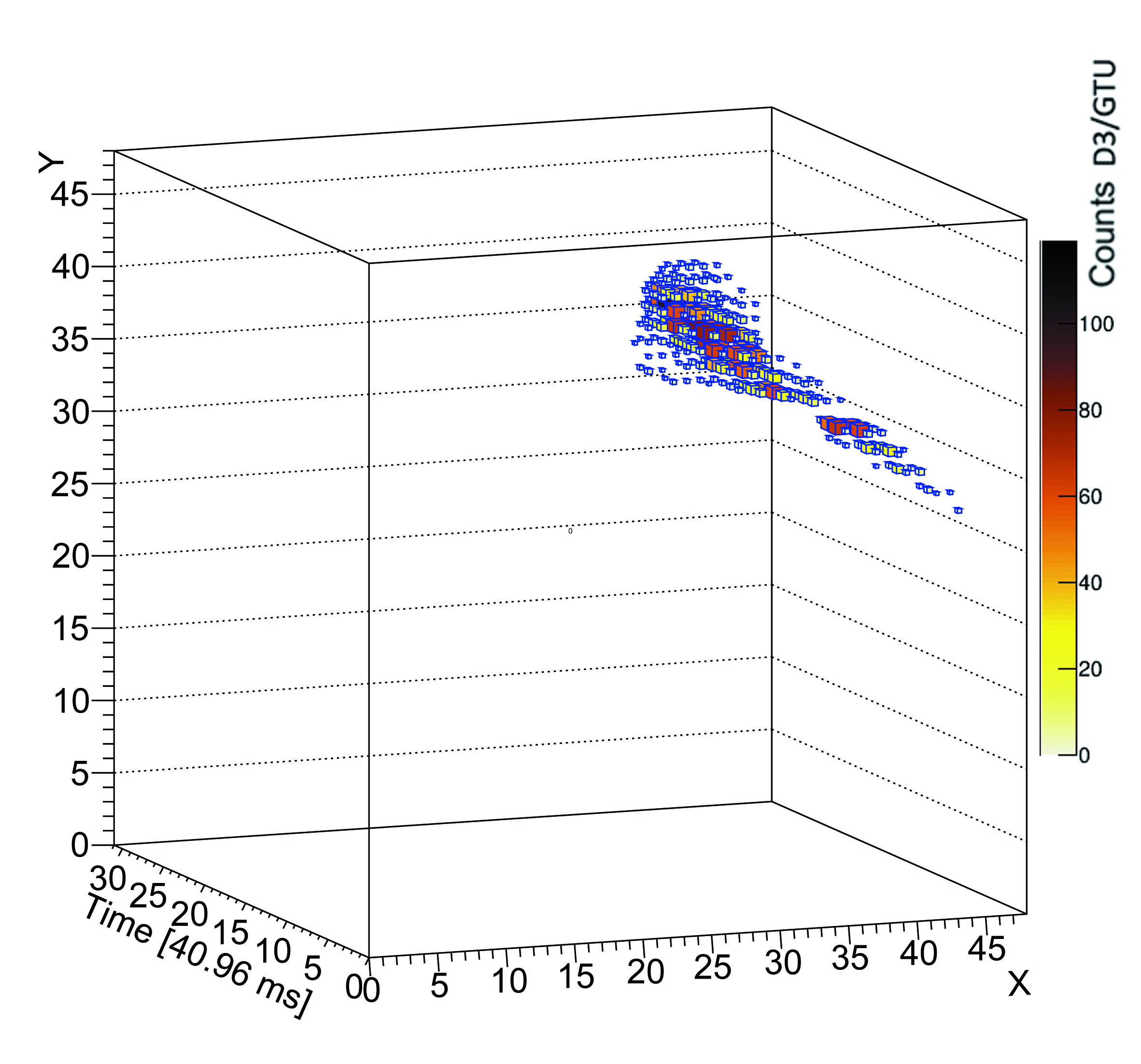}
\includegraphics[width=0.3\textwidth]{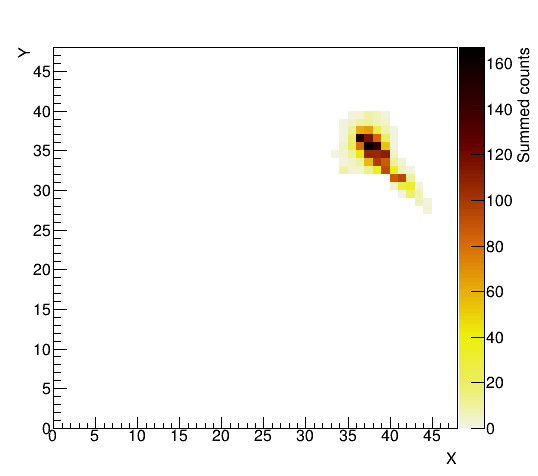}
\includegraphics[width=0.3\textwidth]{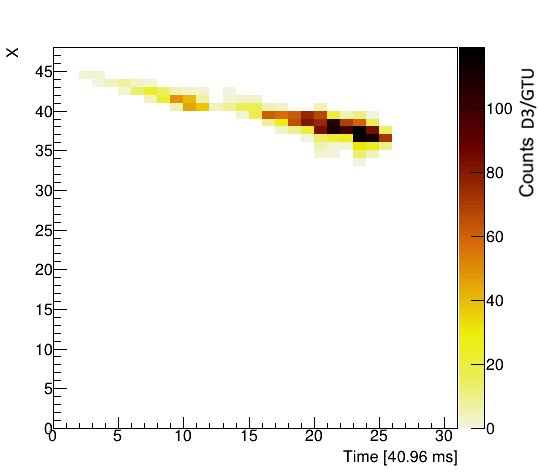}
\includegraphics[width=0.3\textwidth]{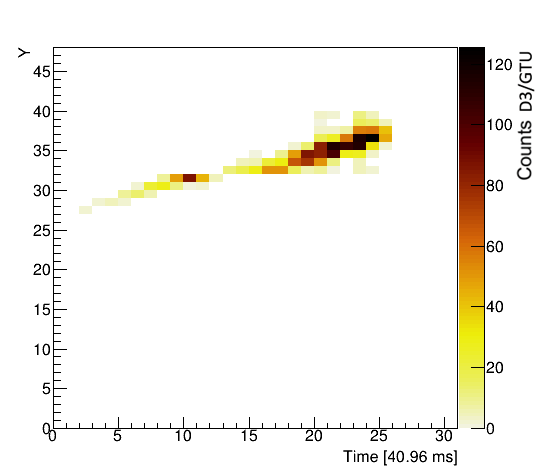}
\caption{Top: A meteor track as it develops in the field of view of Mini-EUSO. The X and Y axis represent the PDM pixels. Bottom: the meteor track projected on the PDM (xy - Left), and on  the x-t and y-t profiles (centre and right, respectively).   Colour denotes counts/GTU (2.5 $\mu$s).}
\label{fig-Meteor}       
\end{figure}
\begin{figure}[ht].
\centering
\includegraphics[scale=0.25]{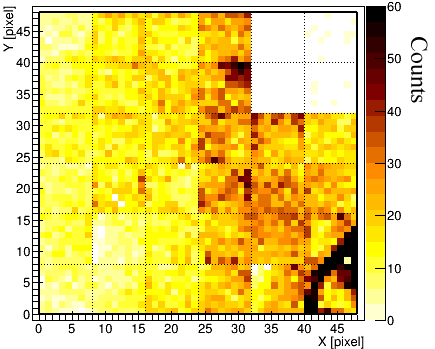}
\includegraphics[scale=0.25]{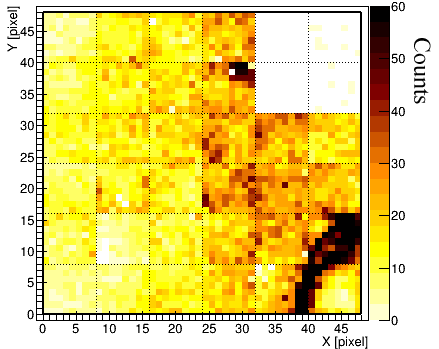}
\includegraphics[scale=0.25]{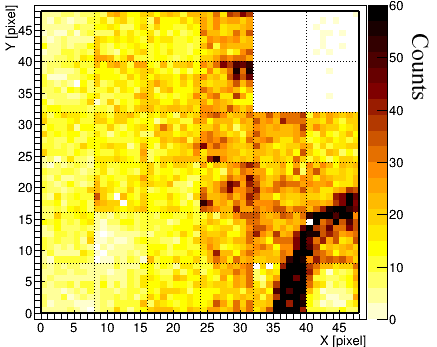}
\includegraphics[scale=0.25]{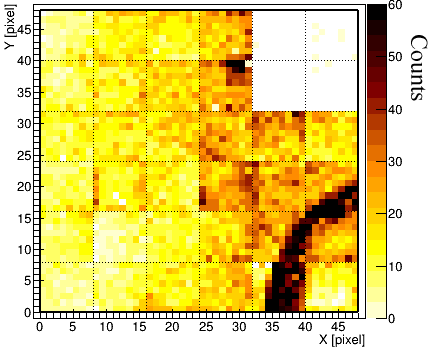}
\includegraphics[scale=0.25]{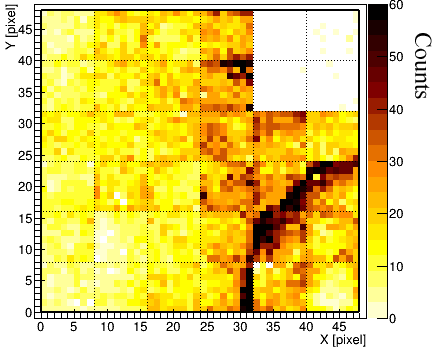}
\includegraphics[scale=0.25]{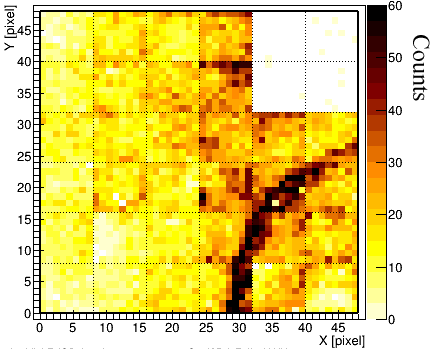}
\includegraphics[scale=0.25]{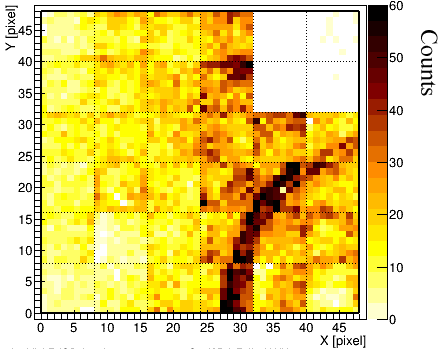}
\caption{A sample of  frames of an Elve being observed in the focal surface. The top right EC unit of four MAPMT is temporarily working at a reduced voltage (about 1/100 sensitivity) due to a previous bright light that triggered the safety mechanism. }
\label{fig-ELVE}        
\end{figure}
\begin{figure}[ht].
\centering
\includegraphics[width=0.8\textwidth]{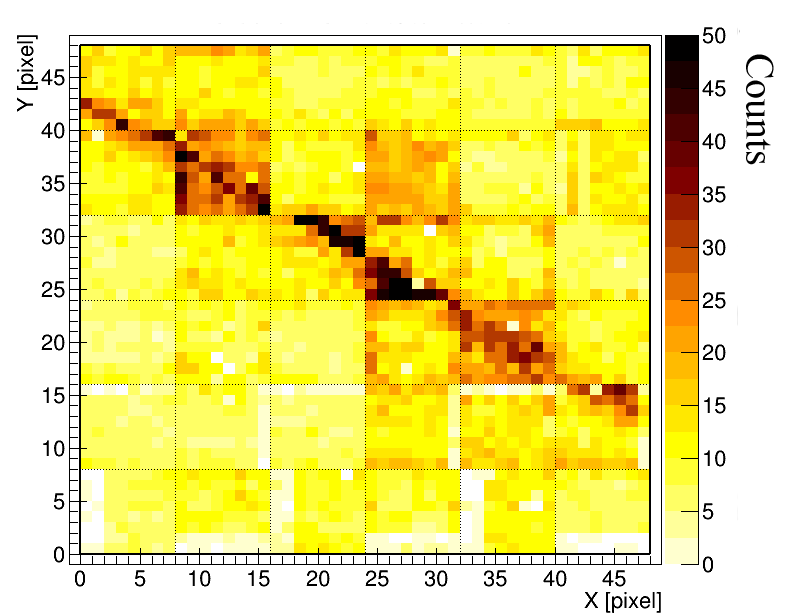}
\caption{One frame of the signal generated by a cosmic ray directly hitting  the focal surface. This is a peculiar event, parallel to the focal surface; whereas most of these events  are more inclined and hit only a few pixels. }
\label{fig-cosmicray}       
\end{figure}

The various signals detected by Mini-EUSO can be distinguished according to their temporal and spatial profile. The main types of  events are:

\begin{enumerate}  
\item{Earth} emissions depend on the surface visible, e.g. ground, sea or clouds. They  move in the field of view with an apparent speed close to the orbital velocity of the ISS\footnote{Since the Earth is also rotating, the apparent speed is slightly lower and closer to 7.4 km/s.}, $\simeq$7.7 km/s. A given point is thus visible for about 42 s ($\simeq 1000$ frames in D3 acquisition) as it moves on the focal surface: it is thus possible to derive ground maps with good spatial resolution and reduced statistical fluctuations. Typical maps of the Earth emissions in the UV are shown in Figures \ref{fig-nighttime} and \ref{frame}.    
Towns  and other anthropogenic lights appear to move at the speed of the ISS. The signal on a given pixel depends  on the size of the town and  its neighbourhood.  The typical time  profile of a single pixel consists of a gradual growth of light lasting for some seconds, according to the size of the town (Figure \ref{overview}).  

\item{Lightning} are transient events with duration of $\simeq$ 1 s and that can illuminate the whole focal surface. As already mentioned, large events can temporarily switch off one or more EC units.  Figure \ref{overview} shows the time profile of some lightning events as  seen in D2 and D3 modes. 
\item{Light modulation}. Artificial lights can also be identified in D2  time scale  in the 50 or 60 Hz light modulation. This is  visible better in small towns and villages which are all connected to the same transformer and which emit light in phase. This is more difficult to observe in larger cities, which have different sections connected to different transformers and with varying  phases.   Figure \ref{overview}   shows the light modulation from Canada and India with 120 and 100 Hz  frequency (double of the AC frequency) respectively. 
\item{Meteors} can have varying signals depending on mass, velocity and angle of incidence. Meteors are identified offline in the D3 time scale looking for straight tracks moving in the field of view. The rate of observed meteors is $\simeq$0.4/minute and it will be the subject of a future paper.  Figure \ref{fig-Meteor} shows the time and spatial profile of a meteor. 
\item{Ground flashers} are   used as obstruction lights (usually with Xenon) to warn aircraft of the presence of   buildings or towers. They have different brightness and duration \citep{ADAMS20141506} but usually last a few hundred $\mu s$ (Figure \ref{overview})  and are usually observed several times as they move  in the field of view of the instrument. 
\item{Elves} are observed as large ring-like upper atmospheric emissions that  appear to be expanding at superluminal speed. In Figure \ref{fig-ELVE} are shown the pictures of one Elve (observed on December 05, 2019) entering the field of view.
\item{Direct hits} on the focal surface are due to cosmic rays that directly interact  with the  photocathode or the BG3 filter either with direct ionization or emitting Cherenkov light.   Most of the events cross one or a few pixels, releasing a high signal that lasts a few GTUs and exhibit a sharp increase and an exponential decrease due to the de-excitation of the elements hit.  Figure \ref{fig-cosmicray} shows a direct hit co-planar to the focal surface  with an exponential decrease lasting about 3-5 frames, depending on the energy deposited by the primary ion. As was to be expected due to the short exposure, the search for UHECR has so far yielded no results.
\end{enumerate}


\section{Conclusions}\label{Conclusions}
In this work we have described the Mini-EUSO telescope and its main characteristics. The detector is currently on board the ISS performing periodic observations of the Earth. Operations are expected to continue for at least three years. Initial analysis of the  data received in the first six months of operations confirm the correct functioning of the instrument and the possibility to fulfill its scientific goals. We have observed events in all the operational time frames, from the fast Elves (2.5 $\mu$s sampling), to meteors (40.96 ms readout), lighting and terrestrial emissions.  Analysis of the data is in progress and will be the subject of future publications.

\acknowledgments
This work was   supported by   State Space Corporation ROSCOSMOS, by the Italian Space Agency through the ASI
INFN agreement n. 2017-8-H.0 and contract n. 2016-1-U.0, by Basic Science Interdisciplinary Research Projects of RIKEN
and JSPS KAKENHI Grant (JP17H02905, JP16H02426 and JP16F16737), by the Italian Ministry
of Foreign Affairs and International Cooperation,  by the French space agency CNES, 
National Science Centre in Poland grant 2017/27/B/ST9/02162.
ST was an International Research Fellow of the Japan Society for the Promotion of Science (2016-2018).

The authors express their deep and collegial thanks to the entire JEM-EUSO program and all its individual members.




\end{document}